# Long-Range Electron Transport in Prussian Blue Analog Nanocrystals.

Roméo Bonnet[1,#], Stéphane Lenfant[1], Sandra Mazérat[2],
Talal Mallah[2],* and Dominique Vuillaume[1],*

1. Institute for Electronics Microelectronics and Nanotechnology (IEMN), CNRS, Av. Poincaré, 59652 Villeneuve d'Ascq, France.
2. Institut de Chimie Moléculaire et des Matériaux d'Orsay (ICMMO), CNRS, Université Paris-Saclay, 91405 Orsay Cedex, France.

# Present address: ITODYS, CNRS, Univ. Paris-Diderot, 15, rue Jean Antoine de Baïf, 75013 Paris, France.
*Corresponding authors : talal.mallah@u-psud.fr; dominique.vuillaume@iemn.fr

**Abstract.** We report electron transport measurements through nano-scale devices consisting of 1 to 3 Prussian blue analog (PBA) nanocrystals connected between two electrodes. We compare two types of cubic nanocrystals, CsCo$^{III}$Fe$^{II}$ (15 nm) and CsNi$^{II}$Cr$^{III}$ (6 nm), deposited on highly oriented pyrolytic graphite and contacted by conducting-AFM. The measured currents show an exponential dependence with the length of the PBA nano-device (up to 45 nm), with low decay factors β, in the range 0.11 - 0.18 nm$^{-1}$ and 0.25 - 0.34 nm$^{-1}$ for the CsCoFe and the CsNiCr nanocrystals, respectively. From the theoretical analysis of the current-voltage curve for the nano-scale device made of a single nanoparticle, we deduce that the electron transport is mediated by the localized d bands at around 0.5 eV from the electrode Fermi energy in the two cases. By comparison with previously reported *ab-initio* calculations, we tentatively identify the involved orbitals as the filled Fe(II)-t$_{2g}$ d band (HOMO) for CsCoFe and the half-filled Ni(II)-e$_g$ d band (SOMO) for CsNiCr. Conductance values measured for multi-nanoparticle nano-scale devices (2 and 3 nanocrystals between the electrodes) are consistent with a multi-step coherent tunneling in the off-resonance regime between adjacent PBAs, a simple model gives a strong coupling (around 0.1 - 0.25 eV) between the adjacent PBA nanocrystals, mediated by electrostatic interactions.

**Introduction.**

Metal Organic Frameworks (MOFs) are porous coordination networks that are intensively explored because they hold tremendous promises for applications. Recently, a focus on conducting MOFs emerged for,[1, 2] beyond the fundamental understanding of their electron transport mechanisms, the design of chemical sensors where information can be electrically read-out.[3] Measurements of the conductance of ultra-thin films (few to 30 nm thickness) of MOFs were carried out on few systems. They show an electron transport behavior characterized by an exponential decay versus distance with a decay factors ($\beta$) in the range 0.4-1 $nm^{-1}$ (with $I \propto e^{-\beta d}$, d being the film thickness), generally larger than those of metal containing molecular wires ($\leq 0.3$ $nm^{-1}$).[4, 5]

Prussian blue analogs (PBAs) and related cyanide bridged systems are also coordination networks that have been known for decades because of their unique optical, magnetic, conducting and electrochemical properties.[6-12] The preparation of nanocrystals of PBA by spontaneous stabilization in water[13, 14] opened tremendous perspectives for applications in a large range of fields,[15] and recently in the domain of energy as battery materials.[11, 12, 16] The reports on the electronic transport properties of PBAs are scarce even at the micro-macroscale (powder, film) level.[17-19] For example, electrical characterization of 50-100 nm thick films of $K\{Fe^{III}[Fe^{II}(CN)_6]\}$ showed non-linear current-voltage curves with a high resistivity suggesting tunneling-limited electron transport between adjacent PBA nanoparticles and/or at the interface with electrodes.[18] Temperature-dependent conductivity measurements of Prussian blue (PB, $Fe^{III}_4[Fe^{II}(CN)_6]_3$) powder films (100 μm thick) and its partially oxidized derivative (also known as Berlin Green, BG) showed hopping transport in BG while PB was found to be insulating.[19] Conductivity measurements on pressed pellet films (hundreds of μm thick) of $Rb_{0.5}\{Co^{III}Co^{II}_{0.25}[Fe^{II}(CN)_6]\}$ and $Rb_{0.8}\{Mn[Fe(CN)_6]_{0.93}\}$ PBAs demonstrated electric field-induced conductance transitions and thermally induced conductivity switching.[10, 20] For ultra-thin films, even though PBA monolayers and few-monolayers have been prepared by various methods such as Langmuir-Blodgett and sequential growth in solution and their structural and magnetic properties studied,[21-23] no electron transport property has been reported so far, albeit this knowledge is of prime importance for the design of PBA-based nanoscale devices.

Here, we report on the electron transport at the nanoscale of devices consisting of 1 to 3 PBA cubic shaped nanocrystals between two electrodes. We compare the electron transport of nanocrystals corresponding to two different PBA networks Cs{Co$^{III}$[Fe$^{II}$(CN)$_6$]} and Cs{Ni$^{II}$[Cr$^{III}$(CN)$_6$]} noted CsCoFe (15 nm) and CsNiCr (6 nm) in the following. The PBA nanocrystals were assembled from solution (see Methods and the Supporting Information) on the surface of highly oriented pyrolytic graphite (HOPG). They form clusters on the HOPG surface with heights from 1 to 3 objects according to scanning electron microscope images and topographic AFM. Current-voltage histograms were measured at room temperature by conducting-AFM (C-AFM). The measured currents show an exponential dependence with the length of the nano-scale devices (up to 45 nm for CsCoFe and 18 nm for CsNiCr), i.e. the number of PBA nanocrystals in the devices, $I \propto e^{-\beta d}$. We deduce low decay factors, $\beta \approx 0.11 - 0.18$ nm$^{-1}$ for CsCoFe, and $\beta \approx 0.25 - 0.34$ nm$^{-1}$ for CsNiCr. For the single PBA nanoscale device, we use a theoretical analysis of the current-voltage curve (single energy level model) and we determine that the orbitals involved in the electron transport are located at around 0.5 eV from the electrode Fermi energy for the two PBA systems. Relying on previously reported *ab initio* calculations,[24-26] we tentatively ascribe the involved orbitals as the filled Fe$^{II}$-t$_{2g}$ d band (HOMO) for CsCoFe, which has been calculated at about 0.2-0.3 eV below the Fermi energy[26] and the half-filled Ni$^{II}$-e$_g$ d band (SOMO) for CsNiCr theoretically predicted at around 0.25 eV above the Fermi energy, the other orbitals being far away.[24-26] Conductance values measured for multi-nanoparticle devices (2 and 3 nanocrystals between the electrodes) were analyzed with a multi-step coherent off-resonance tunneling between adjacent nanocrystals, a simple model gives a strong coupling (around 0.1 – 0.25 eV) between the nanoobjects, likely due to Cs$^+$ mediated electrostatic interactions.

**Results.**

PBAs have a face centered cubic (fcc) structure made by a trivalent (in most cases) transition metal ion linked to a divalent metal ion through the cyanide bridge with a cell parameter close to 10 Å (Scheme 1-a). The tetrahedral sites of the fcc structure may contain alkali ions (A, Cs here) controlling the concentration in $[M(CN)_6]^{n-}$ vacancies throughout the cubic network. The chemical composition (nature of the metal ions, concentration of alkali ions and $[M(CN)_6]^{n-}$ vacancies) can be finely tuned leading to the different physical and chemical properties characterizing the PBAs mentioned above.

Here, we focus on the $Cs\{Co^{III}[Fe^{II}(CN)_6]\}$ and $Cs\{Ni^{II}[Cr^{III}(CN)_6]\}$ PBA nanocrystals (CsCoFe and CsNiCr, respectively). CsCoFe is made from diamagnetic low spin ($S = 0$, $t_{2g}^6 e_g^0$) $Co^{III}$ and $Fe^{II}$ metal ions. While CsNiCr has paramagnetic $Ni^{II}$ ($S = 1$, $t_{2g}^6 e_g^2$) and $Cr^{III}$ ($S = 3/2$, $t_{2g}^3 e_g^0$). We and others have already demonstrated that the nanoscale dimension brings additional properties to these coordination networks. For example, the CoFe network in the powder form presents an electron transfer at low temperature upon light irradiation leading to the magnetic $Co^{II}$ ($S = 3/2$, $t_{2g}^5 e_g^2$)-$Fe^{III}$ ($S = ½$, $t_{2g}^5 e_g^0$) state.[8, 9] The CsCoFe nanocrystals have the remarkable property to undergo a fast (ps range) electron transfer at room temperature after light irradiation.[8, 27, 28] The CsNiCr nanocrystals display a magnetization reversal,[29, 30] that was shown to occur at the level of a single object.[31] Recently, 8 nm CsNiCr nanocrystals were coupled to resonant microwave fields opening the route for their use as components for quantum information technology.[32] These nano-objects are at the frontier between bulk crystalline materials and molecules, because they retain the properties of the bulky materials but they can be processed in solution to form isolated nanoobjects, monolayers and few multilayers on surfaces (like molecules) and can, therefore, be used for different applications.[15] We, therefore, took advantage from their stabilization in water without any surfactant to assemble them on HOPG and measure their electron transport properties in devices made of one, two and three nanocrystals.

The CsCoFe and CsNiCr PBA nanocrystals were prepared as previously reported in Refs.[13, 28] and the main steps are summarized in Methods and detailed in the Supporting Information, as well as their physicochemical characterization. Dynamic light scattering (DLS) shows a hydrodynamic diameter of 6 nm (for CsNiCr, Fig. S1 in the Supporting Information) and 15 nm (for CsCoFe, Fig. S5 in the Supporting Information), a size confirmed by transmission electronic microscopy (TEM) (Figs. S2 and S6 in the Supporting Information). Infra-red spectroscopy shows the characteristic peaks of the bridging cyanide assigned to $Cr^{III}$-CN-$Ni^{II}$ (2171 $cm^{-1}$) and $Fe^{II}$-CN-$Co^{III}$ (2120 $cm^{-1}$) sequences, respectively (Figs S3 and S7 in the Supporting Information). X-ray powder diffraction (Figs. S4 and S8 in the Supporting Information) are consistent with nanocrystals having a face-centered cubic (fcc) structure (Scheme 1-a) and a unit cell parameter close to 10 Å as expected.[9, 29] The size of the crystalline domains (6.2 and 11 nm for CsNiCr and CsCoFe, respectively) confirm the DLS and TEM results. Finally, energy dispersive X-ray spectroscopy (see the Supporting Information) indicates the following compositions $Cs_{0.96}Ni[Cr(CN)_6]_{0.94}$ and $Cs_{0.7}Co[Fe(CN)_6]_{0.9}$. In summary, scheme 1-a gives a view of the unit cell of the crystalline PBA nanocrystals. Part of the $Cs^+$ ions occupy the tetrahedral sites and because the nanocrystals are negatively charged, they also play the role of counter cations.[13, 14] Finally, the surface of the nanocrystals have both terminal water molecules and nitrogen cyanide atoms.[33] The water molecules are coordinated to the Ni(II) (or Co(III)) ions present at nanocrsytals' surface and the nitrogen atoms belong to the $M'(CN)_6$ surface species (scheme 1-b).

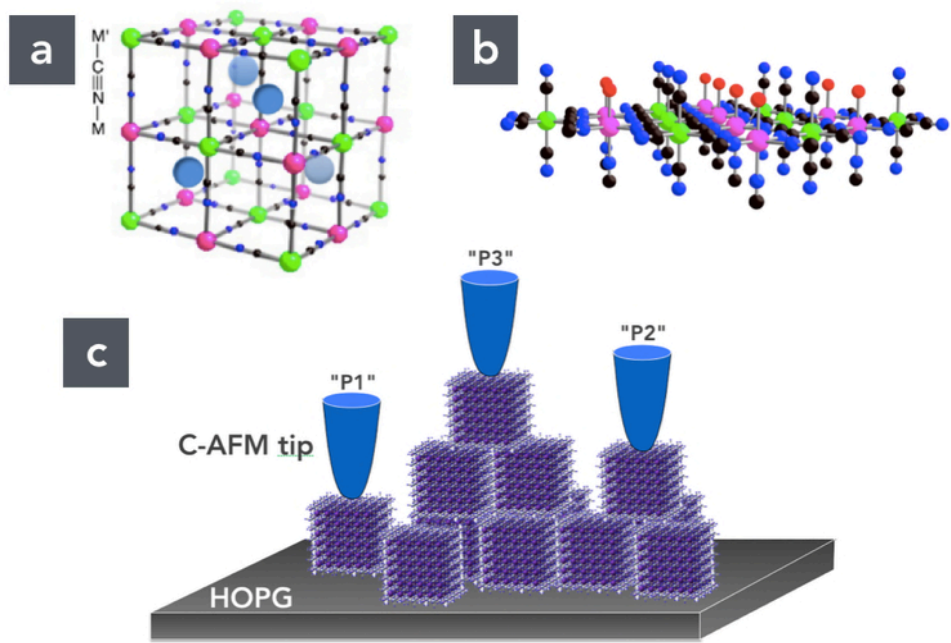

**Scheme 1**. *(a) View of the unit cell of the fcc structure of the PBA nanocrystals, where M' = Cr(III) or Fe(II) (green) and M = Ni(II) or Co(III) (purple), the tetrahedral sites are occupied by the Cs$^+$ ions (blue). (b) View of the nanocrystals' surface with oxygen atoms (red) belonging to water molecules. (c) Schematic view of the HOPG/PBAs/C-AFM tip nano-scale devices with 1, 2 and 3 PBA nanocrystals.*

Figure 1 shows typical scanning electron microscope (SEM) images and atomic force microscope images (AFM) of the CsCoFe and CsNiCr nanocrystals deposited on highly oriented pyrolytic graphite (HOPG) substrates from a colloidal solution of the as-prepared nanocrystals (see Methods). The SEM images clearly show that the conditions used for the deposition produce a partially covered surface with different structures (clusters) of the deposited nanocrystals. A more detailed analysis of the AFM images (histograms of heights, Figs. 1-d and 1-f) confirms the presence of monolayers and multilayers of nanocrystals by comparison with their measured nominal size (see the Supporting Information, nominal values: s = 15 nm for the CsCoFe, s = 6 nm for the CsNiCr).[29, 30] The histograms show peaks that are multiples of the nominal sizes of the nanocrystals, with standard deviations comparable with those measured by TEM on as-synthesized objects. We note that we have never observed more than 3 layers of nanocrystals with the deposition conditions (see Methods) used in this study (see additional AFM images and histograms of heights, Figs. S9 and S10 in the Supporting Information). We also analyzed the distribution of the particles' sizes (grain analysis) from the AFM and SEM images (Fig. S11 in the Supporting Information) and we observed a maximum of counts centered at ca. 15 nm and ca. 6 nm for the CsCoFe and CsNiCr nanocrystals, respectively, with some domains of larger sizes due to aggregation. We note that this morphological information is in good agreement with the TEM and DLS characterization (see the Supporting Information) and already published results on CsNiCr.[13, 30]

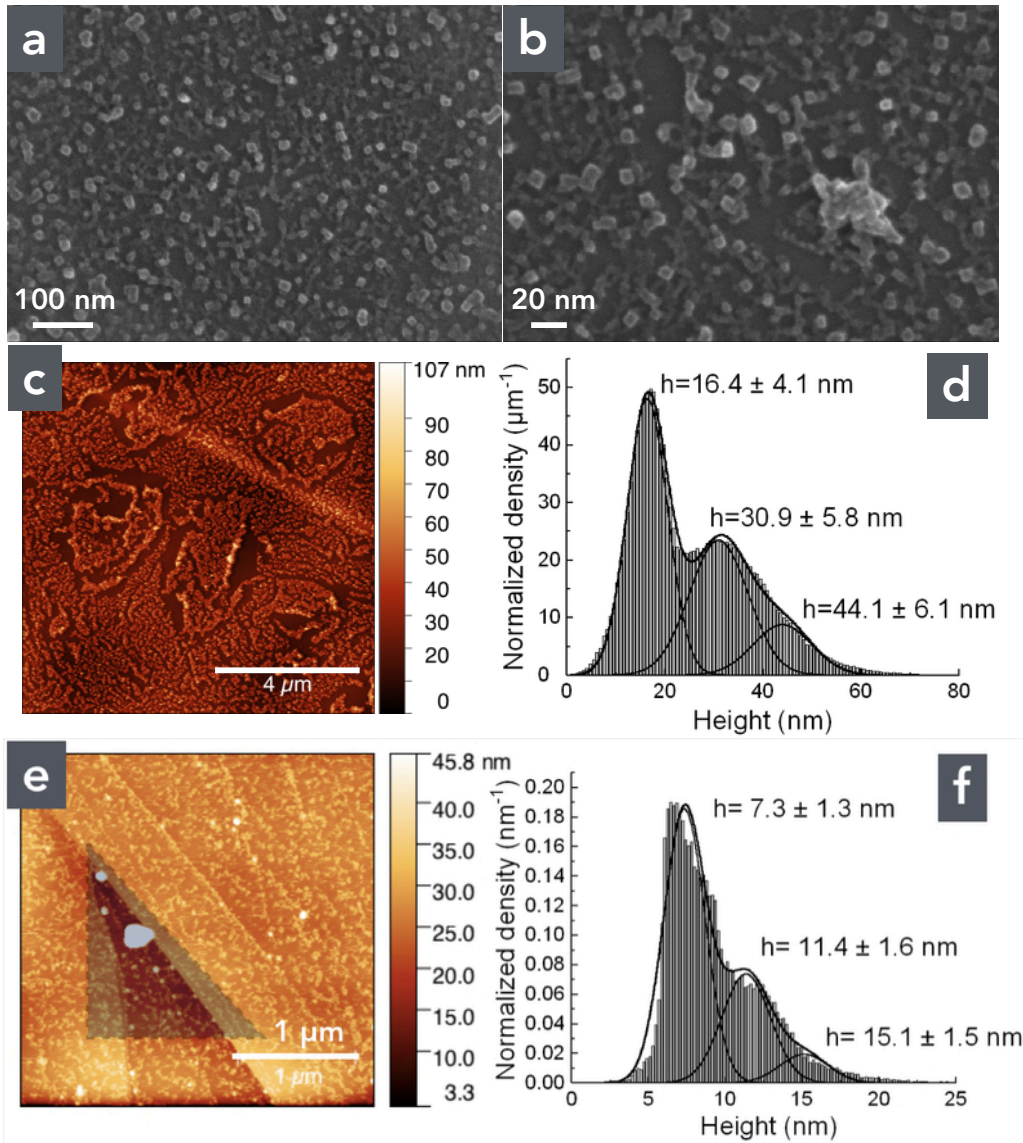

*Figure 1. SEM images (7kV, secondary electron images) of the CsCoFe layers on HOPG at two magnifications: **(a)** x 141.6k, **(b)** x 221.7k. **(c, d)** topographic TM-AFM image and corresponding histograms of heights of the CsCoFe layers. **(e, f)** topographic TM-AFM image and corresponding histograms of heights of the CsNiCr layers. The shadowed triangle indicates a region discarded from the histogram analysis to avoid the contribution of a thick "contamination" (white spot, height about 150 nm, and the triangular hollow (about 10 nm deep). The other light steps (more or less parallel in the image have a weak height (about 1*

*nm) and they are likely coming from the substrate (the corresponding weak peak was removed in the histogram). In panels (d) and (f), the values marked for each peak are the average heights and the standard deviations obtained from the fits of multi-Gaussian distributions (black lines). Additional AFM images and histograms of heights, Figs. S9 and S10 in the Supporting Information.*

Figures 2 and 3 show the current-voltage (I-V) 2D histograms for the CsCoFe and CsNiCr nanocrystals, respectively, measured by C-AFM (scheme 1-c). These 2D histograms reveal largely dispersed values of the currents, which are distributed in several groups of I-V curves. To further analyze these current distributions, 1D current histograms were extracted at 400, 200 and 50 mV and fitted by several log-normal laws. For the CsCoFe nanocrystals, we identify 2 main peaks (P0 and P1, Fig. 2a) from the highest measured currents, and two peaks (P2 and P3, Fig. 2b) for the lowest measured currents. The same analysis for the CsNiCr nanocrystals gives 4 peaks (P0-P3) for the currents measured with a low amplifier sensitivity (Fig. 3a) and high amplifier sensitivity (Fig. 3b) (see Methods for the measurement details). For each peak of the current distribution, a log-mean current, log-$\mu$, and a log-standard deviation, log-$\sigma$, are deduced from the fits shown in Figs. 2 and 3. The values are reported in Table 1.

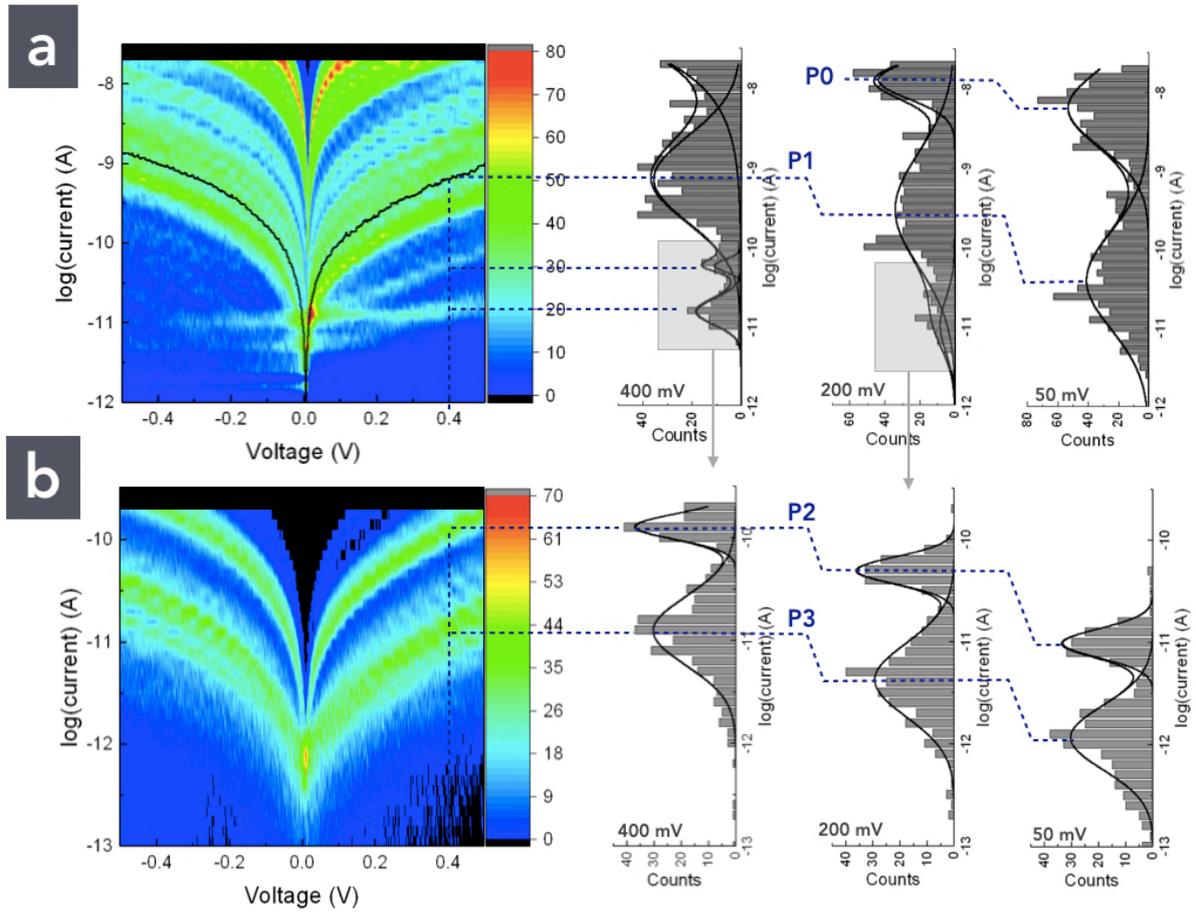

*Figure 2. CsCoFe. C-AFM 2D current-voltage (I-V) histograms and corresponding 1D histograms at 400mV, 200mV and 50mV (all currents plotted on decimal log scales): **(a)** currents measured at a low amplifier sensitivity (500 I-V traces) and **(b)** currents measured at a higher amplifier sensitivity (500 I-V traces). The blue dashed lines are guide for eyes to identify the different peaks in the 1D current histograms. The black lines are fits with log-normal distributions, the log-mean current, log-µ, and log-standard deviation, log-σ, are given in table 1. The grey areas mean that the fit parameters are not taken from these measurements but from the current measurements with a higher amplifier sensitivity for a better resolution. The black line in the 2D I-V histogram is the average I-V for peak P1 (see text).*

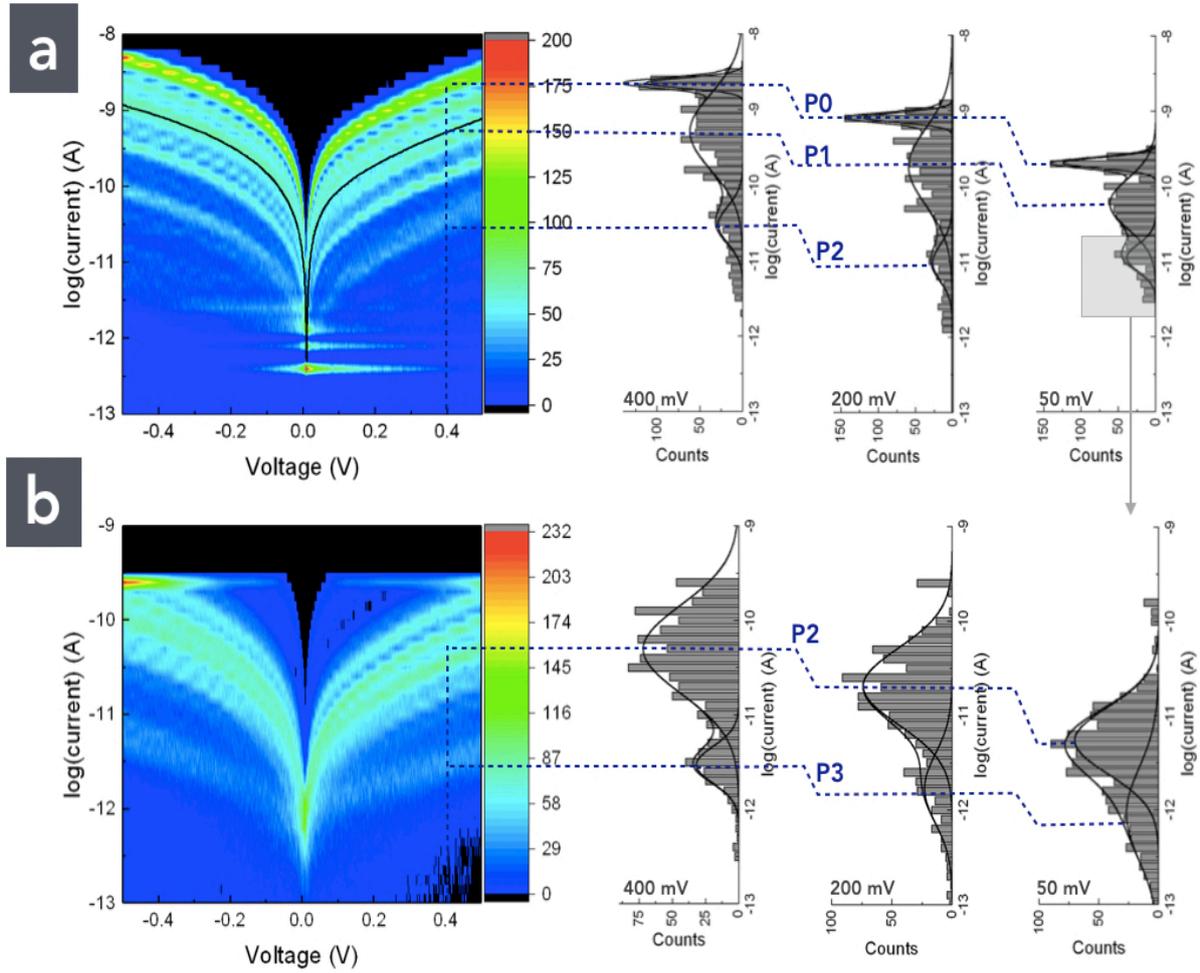

*Figure 3.* CsNiCr. C-AFM 2D current-voltage (I-V) histograms and corresponding 1D histograms at 400mV, 200mV and 50mV (all currents plotted on decimal log scales): **(a)** currents measured at a low amplifier sensitivity (500 I-V traces) and **(b)** currents measured at a higher amplifier sensitivity (500 I-V traces). The blue dashed lines are guide for eyes to identify the different peaks in the 1D current histograms. The black lines are fits with log-normal distributions, the log-mean current, log-µ, and log-standard deviation, log-σ, are given in table 1. The grey area means that the fit parameters are not taken from these measurements but from the current measurements with a higher amplifier sensitivity for a better resolution. The black line in the 2D I-V histogram is the average I-V for peak P1 (see text).

Considering the measured topography of the samples (Fig. 1), we ascribe each current peak to electron transport through HOPG/PBA/C-AFM tip with different thicknesses of the PBA layer. By comparison with C-AFM measurements on the same bare HOPG substrate (Figure S12 in Supporting Information) and considering the incomplete coverage of the surface (Fig. 1), the peak P0 is ascribed to HOPG. Then, again considering the topographic AFM images (Fig. 1) showing a stack of the PBA nanocrystals, we assume that P1, P2 and P3 correspond to one, two and three monolayers of PBA nanocrystals, respectively (scheme 1-c). We tried to correlate the scanning topographic images and the scanning current images (simultaneously recorded during the C-AFM measurements in the scanning mode) to identify the current peaks with specific PBA layer heights, but because these measurements required using the contact mode and even with a low loading force (ca. 3 nN, see Methods), we have observed a tendency of a deformation of the PBA layer topography compared to the soft TM-AFM presented in Fig. 1. Thus, we only used a non-scanning approach measuring I-V curves at random points on the surface (see Methods). In addition to the current histograms (Figs. 2 and 3), this method also records the tip z-position during the I-V measurements, and we constructed the corresponding tip z-position histograms (at 0.4 V for comparison with current histograms), the z-position being constant during a given I-V measurement). On these z-position histograms, we also observe (Figs. S12 and S13 in the Supporting Information) 4 peaks spaced on average by 17.3 and 7.9 nm for the CsCoFe and CsNiCr samples, respectively, in reasonable agreement (considering the standard deviation of a few nm) with the known heights of these PBA nanocrystals ($s = 15$ nm for CsCoFe, $s = 6$ nm for CsNiCr) in agreement with the measured height profiles by AFM (Fig. 1). Thus, we conclude that these peaks correspond to the P0-P3 peaks observed from the current histograms (P0 = substrate, P1 = 1 PBA layer, P2 = 2 PBA layers and P3 = 3 PBA layers). Despite the fact that our measurement method does not allow relating directly current and topography, the agreement between current and z-position histograms corroborate our interpretation of the peaks. Figure 4 presents the thickness dependence of the mean current peaks, (log-µ) vs. d, d being the PBA layer heights measured from the height histograms in Fig. 1. We obtain an exponential thickness dependence of the current, I ∝ $e^{-\beta d}$, with a decay factor β ≈ 0.11 - 0.18 $nm^{-1}$ for the CsCoFe nanocrystals, and β ≈ 0.25 - 0.34 $nm^{-1}$ for the CsNiCr nanocrystals. These factors are independent from the applied voltage. Albeit, we note a large difference of the current values of the P0 peak for the two samples (HOPG substrate), the data measured on the PBAs are reasonably extrapolated (Fig. 4) to their corresponding HOPG substrates (except for CsCoFe at 400 mV since the peak P0 is located above the saturation of the instrument, see Fig. 2-a). This difference in the HOPG currents (around $10^{-8}$ A for the CsCoFe samples and $10^{-9}$-$10^{-10}$ A for the CsNiCr samples) is rationalized because it is known that bare HOPG have a large dispersion of conductance depending on the exact sheets, ribbons, step edges contacted by the C-AFM tip with currents measured from tens of nA (at low bias < 1 V) to µA and larger.[34] This behavior was also observed for our HOPG substrates (figure S14 in Supporting Information). Since the I-V traces are recorded in a "blind" mode (see Methods) without a choice of the exact location on the sample surface, this difference of the P0 values reflects this known dispersion. This means that the I-V curves for the CsCoFe samples have been measured on a relatively high conducting HOPG substrate (as in Figs. S14-a and b in the Supporting Information), while the I-V curves for the CsNiCr were acquired on a less conducting HOPG zone (also on another substrate) as shown in Fig. S14-c (Supporting Information). Moreover, the large distribution of the current (Figs. 2 and 3) may be partly due to the conductivity variations of the underlined HOPG, but also from the fact that some PBAs are laterally "connected" (see Figs. 1 and S11) which tends to modulate the number of conducting pathways between the C-AFM tip and the substrate. We also note that we have not observed stacks with more than 3 PBAs in these samples prepared from the colloidal solutions (see Methods) as discussed above from AFM analysis (Fig. 1). Nevertheless, we also prepared samples with a denser and more compact PBA film with a complete coverage of the HOPG surface (longer immersion time of 15 min, see Fig. S15 in the Supporting Information), but no current was detected in agreement with the extrapolation of a current below 10$^{-12}$ A for more than 3 PBA layers in Fig. 4.

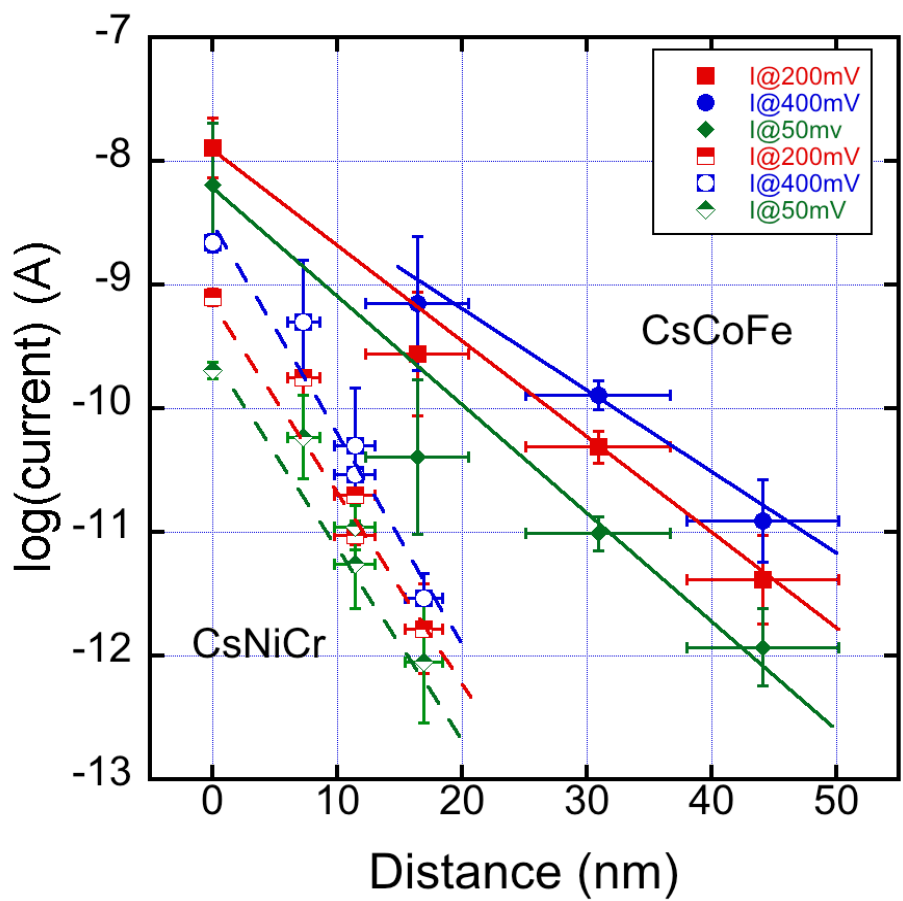

*Figure 4.* *Thickness dependence of the current peaks (log-mean current, log-µ, from 1D histograms in Figs. 2 and 3) for the HOPG/PBAs/C-AFM tip junctions (all currents plotted on decimal log scales). The y-scale error bars are the fitted log-standard deviation, log-σ (see table 1). The x-scale values and error bars are taken from the histograms of heights (average and standard deviation) in Fig. 1.*

|  | P0 | | | P1 | | | P2 | | | P3 | | |
| --- | --- | --- | --- | --- | --- | --- | --- | --- | --- | --- | --- | --- |
|  | 400 mV | 200 mV | 50 mV | 400 mV | 200 mV | 50 mV | 400 mV | 200 mV | 50 mV | 400 mV | 200 mV | 50 mV |
| **CsCoFe** | | | | | | | | | | | | |
| log-µ | n.m. | -7.89 | -8.19 | -9.14 | -9.57 | -10.39 | -9.89 | -10.31 | -11.01 | -10.91 | -11.38 | -11.93 |
| log-σ | n.m. | 0.24 | 0.50 | 0.55 | 0.51 | 0.49 | 0.12 | 0.13 | 0.14 | 0.33 | 0.36 | 0.31 |
| **CsNiCr** | | | | | | | | | | | | |
| log-µ | -8.65 | -9.09 | -9.69 | -9.31 | -9.74 | -10.23 | -10.53 / -10.31 | -11.02 / -10.69 | -11.26 | -11.53 | -11.77 | -12.05 |
| log-σ | 0.07 | 0.06 | 0.07 | 0.50 | 0.50 | 0.33 | 0.24 / 0.46 | 0.24 / 0.40 | 0.3 | 0.19 | 0.36 | 0.49 |

*Table 1. Fitted parameters: log-mean current, log-µ, and log-standard deviation, log-σ for the fits of the log-normal distributions shown in Figs. 2 and 3. (n.m. stands for not measurable).*

The decay factor β determined from these averaged data (Fig. 4) is confirmed by a more detailed statistical analysis directly plotting log(I) taken at 400 mV, 200 mV and 50 mV from individual I-V trace in the data sets shown in Figs. 2 and 3 versus the HOPG-tip distance calculated from the tip z-position recorded for each I-V trace (see details in the Supporting Information and results shown in Fig. S16).

**Discussion.**

From the C-AFM measurements on the PBA monolayers (peak P1) we can estimate the conductance of a single PBA nanocrystal. The C-AFM contact area is estimated as S = πa$^2$ ≈ 5-10 nm$^2$ (with $a = \sqrt{R\delta}$, where R the C-AFM tip radius ≈ 30 nm, δ the typical indentation depth on molecular film assumed to be ≈ 0.05-0.1 nm).[35, 36] Albeit the estimation of the contact area is crude, it is sufficient to validate that the measured conductance from the peak P1 (Figs. 2 and 3) gives the conductance of a single PBA nanocrystal because the contact area is well smaller than the area of a face of one cubic nanocrystal (over 200 nm$^2$ for CsCoFe and 36 nm$^2$ for CsNiCr). Thus, considering the measured conductance of the substrate (peak P0), we deduce the mean conductance $G_{CsCoFe}$ ≈ 1.7 x 10$^{-5}$ $G_0$ and $G_{CsNiCr}$ ≈ 3 x $10^{-5}$ G$_0$ (from the I-V slope around 0 V ± 50 mV, G$_0$ is the conductance quantum G$_0$ = 77.5 nS). We also used a single energy level model to fit the average I-V curve of the peak P1 (scheme 2).[37, 38] We determine the energy position of the orbital involved in the electron transport $\varepsilon_0$ (with respect to the average Fermi energy of the electrodes) and the nanocrystal/electrode coupling parameters $\Gamma_1$ and $\Gamma_2$ using the equation

$$I = \frac{8e}{h} \frac{\Gamma_1 \Gamma_2}{\Gamma_1 + \Gamma_2} \left[ \arctan \frac{2\varepsilon_0 + eV \frac{\Gamma_1 - \Gamma_2}{\Gamma_1 + \Gamma_2} + eV}{2(\Gamma_1 + \Gamma_2)} - \arctan \frac{2\varepsilon_0 + eV \frac{\Gamma_1 - \Gamma_2}{\Gamma_1 + \Gamma_2} - eV}{2(\Gamma_1 + \Gamma_2)} \right]$$

(1)

with *e* the electron charge and *h* the Planck constant.

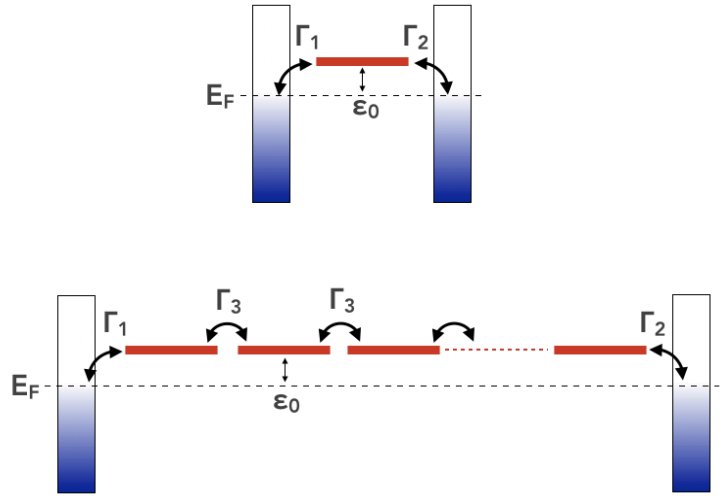

***Scheme 2.*** *Schematic energy diagrams for electron transport though a single PBA and multi PBAs (arbitrarily drawn for example considering the LUMO). $\varepsilon_0$ is the orbital energy with respect to the average Fermi energy of the electrodes and the PBA/electrode coupling are described by coupling parameters $\Gamma_1$ and $\Gamma_2$. Multistep charge carrier transport between adjacent PBAs is parameterized by a third coupling parameter $\Gamma_3$ (Refs. [37-39])*

Figure 5 shows the average I-V traces belonging to the peak P1 (bold black lines in Figs. 2-a and 3-a, we averaged all curves within the FWHM of the log-normal distribution) for both samples and the fits of Eq. 1 with the following parameters: $\varepsilon_0$ = 0.48 eV, $\Gamma_1$ = 1.4 meV and $\Gamma_2$ = 0.9 meV for CsCoFe, $\varepsilon_0$ = 0.51 eV, $\Gamma_1$ = 1.5 meV and $\Gamma_2$ = 0.95 meV for CsNiCr. Not surprisingly, the fitted parameters are close, as the conductance of the single nanocrystals are. Considering the data dispersion (see Figs. 2 and 3), in order to estimate the confidence limits of these parameters, we also fitted two additional I-V curves bounding the peak P1 at the lower and upper limit of the Gaussian distribution (see Figs. S17-a and S17-c in the Supporting Information). We obtain the following values (Figs S17-b and S17-d): $\varepsilon_0$ = 0.42 – 0.55 eV, $\Gamma_1$ = 1 – 1.6 meV and $\Gamma_2$ = 0.57 – 1.4 meV for CsCoFe, $\varepsilon_0$ = 0.43 – 0.54 eV, $\Gamma_1$ = 0.79 – 1.7 meV and $\Gamma_2$ = 0.58 – 1.1 meV for CsNiCr. In a more detailed statistical analysis, we also determine these energy values by fitting the single energy level model on individual I-V trace in the two data sets shown in Figs. 2 and 3 (except discarding several I-V curves with instabilities and large noise given poor fits, see the Supporting Information, Figs. S18 and S19, for CsCoFe and CsNiCr, respectively). We obtain energy values well fitted by a Gaussian distribution: $\varepsilon_0$ = 0.48 ± 0.06 eV (CsCoFe) and $\varepsilon_0$ = 0.53 ± 0.07 eV (CsNiCr). We note that these energy levels are almost the same for all the three peaks P1, P2 and P3. This feature is discussed below (see also Figs. S22 and S23 in the Supporting Information).

We used already reported *ab initio* calculations on the two type of networks, that showed a rather large HOMO-LUMO gap (1.5-2 eV) and localized d bands near the Fermi energy.[24-26] However, the energetic schemes are different for the two cases. For CsCoFe, the first-principles relativistic many-electron calculations[26] showed that the LUMO is located at about 1.6-1.9 eV from the Fermi energy and corresponds to the $e_g$ orbital for Co$^{III}$, while the HOMO is very close to the Fermi energy (ca. 0.2-0.3 eV below it) and it is ascribed to the $t_{2g}$ orbital of Fe$^{II}$. Thus, we can reasonably assume that the molecular orbital involved in the electron transport in the HOPG/CsCoFe/C-AFM tip, measured at $\varepsilon_0$ = 0.42 - 0.55 eV is the Fe$^{II}$-$t_{2g}$ d band (HOMO), the LUMO being far away (scheme 3). On the contrary, the calculated electronic structure of CsNiCr within the formalism of local-spin-density approximation[24] and *ab-initio* DFT calculations,[25] the HOMO at ca. 1 eV below the Fermi energy (due to the band overlapping of ($Ni^{II}$-$e_g$)' ($Ni^{II}$-$t_{2g}$) and ($Cr^{III}$-$t_{2g}$) orbitals) and a half-filled state (SOMO) near the Fermi energy (ca. 0.25 eV) related to the ($Ni^{II}$-$e_g$) orbital. Thus, the SOMO is the nearest level to the electrode Fermi energy and we, reasonably, assume that this SOMO corresponds to the orbital experimentally detected at $\varepsilon_0$ = 0.43 - 0.54 eV in our experiments (scheme 3). This level half-filled level should be expected in resonance with the Fermi energy of the electrodes. However, as recently reviewed,[40] despite significant progress in the theoretical description of the electron transport in molecular junctions, the theory-experiment comparison remains largely qualitative, except in rare cases, e.g. for small molecules and experiments in well-controlled conditions. Here, these comparisons with calculated energy levels of the PBA nanocrystal in gas phase do not consider any charge transfer and interaction between the nanocrystal and the electrodes that likely occur in a solid-state HOPG/PBA/metal junction and that can shift the nanocrystals' energy levels with respect of the Fermi energy of electrodes. More precise energy level identification in the HOPG/PBA/metal will require further more elaborated calculations of the electron transport properties of these PBA-based nano-devices.

Similarly, the weak asymmetry in the I-V curves (a ratio ca. 1.5 between the current at - 0.5V and 0.5 V, as well as the weak difference between the fitted $\Gamma_1$ and $\Gamma_2$, see above), is not significant and cannot be attributed to the nano-object behavior (i.e. HOMO vs. LUMO(or SOMO) dominated transport). A ratio larger than 10 is now admitted as a statistically relevant criterion to claim an electrical rectification in molecular junctions and to be physically interpretable.[41] In principle, such a negative rectification (more current at negative bias) may come from the work function difference between the two electrodes (ca. 4.6 eV and 5.1 eV for HOPG and PtIr AFM tip) with a higher current when a negative bias is applied on the electrode with the lower work function (here HOPG, see Methods), but, again, the present ratio is too small to conclude. Moreover, the fact that $\Gamma_1 \approx \Gamma_2$ is also consistent with measurements on other molecular junctions (small molecules like alkyl chains or π-conjugated oligomers) using asymmetric electrodes (carbon-based and metal, e.g. graphene and Au or graphene and eGaIn). No strong asymmetry (i.e. $\Gamma_1 \approx \Gamma_2$) is induced in the I-V curves in these molecular junctions with asymmetric electrodes compared to the case of the same molecules with symmetric electrodes (Au-molecules-Au).[42-44] Another issue is that the HOPG substrate is not a perfect metal and the exact density of states is not taken into account in the very simple single energy level model (Eq.1). This may explain the slight deviation (voltage sensitivity) from the model at positive bias (Fig. 5, Figs. S17 and S23 in the Supporting Information). Finally, given the large dispersion in the experimental results, better-controlled experiments are required as suggested in the conclusion (e.g. on single PBA nanoscrystals with size controlled in the range 6 – 80 nm,[14, 29] instead on changing the device size by varying the number of involved nanocrystals). At this stage, the present quantitative agreement seems reasonable.

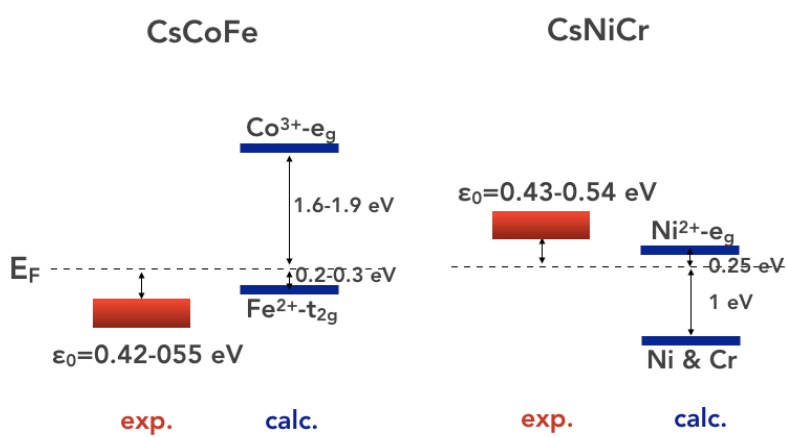

**Scheme 3.** *Energy diagrams for the two PBAs from transport experiments and ab-initio calculations (Refs. [24-26]).*

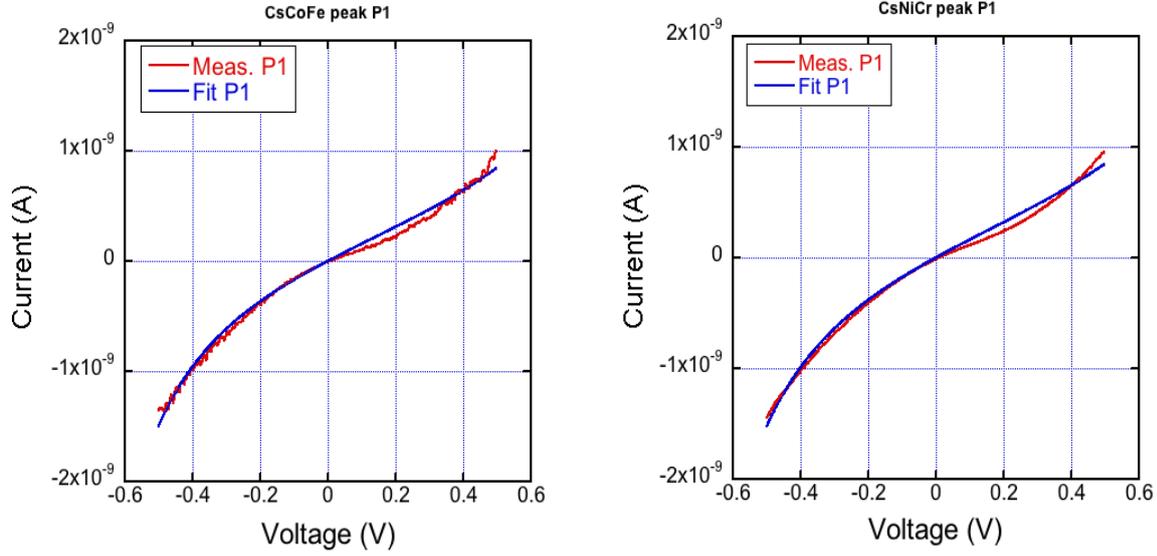

*Figure 5.* Average current-voltage curves of peak P1 (red lines) from the 2D-histograms (black lines in Figs. 2a and 3a) and fit of the single molecular energy level (Eq. 1, blue lines) with: $\varepsilon_0$ = 0.48 eV, $\Gamma_1$ = 1.4 meV and $\Gamma_2$ = 0.9 meV for CsCoFe, $\varepsilon_0$ = 0.51 eV, $\Gamma_1$ = 1.5 meV and $\Gamma_2$ = 0.95 meV for CsNiCr.

For the peaks P2 and P3 involving 2 and 3 nanocrystals, we consider a multistep charge carrier transport between adjacent sites, parameterized by a third coupling parameter $\Gamma_3$ (see scheme 2). We assume that all the nanocrystals in the junction have the same energy level $\varepsilon_0$. Albeit this simplification, this approximation was satisfactorily used to analyze electron transport measurements in metal-coordinated molecular wires.[4, 45] If we consider fluctuations of this energy level[38] from one nanocrystal to another (e.g. nanocrystals in contact with the electrodes are influenced by electrode coupling, while the ones in the center are not) this feature will induce some broadening of the parameter $\Gamma_3$ and/or significant variation of the energy level $\varepsilon_0$ extracted from the I-V measurements on devices with various numbers of nanocrystals (which is not the case, see below the discussion about the similar values of $\varepsilon_0$ for peaks P1, P2 and P3 and Figs. S18, S19, S22 and S23 in the Supporting Information). Considering the experimental dispersion of our measurements, we assume that this simplification is justified here. A standard electron transport mechanism is incoherent hopping between adjacent sites, as observed in long molecular wires (e.g. in π-conjugated molecules up to ca. 40 nm,[4, 46-48] in DNA,[49] see reviews in Refs [5, 38, 39]). However, this mechanism implies that the conductance decreases linearly with the number of sites, i.e as $1/d$.[38, 39] This is clearly not the case here (figure S20 in the Supporting Information) as shown when plotting the conductance versus $1/d$. Plotting the log of the current versus the electric field (figure S21 in the Supporting Information) shows that the transport is not field driven as it should be for a hopping mechanism and as observed in long molecular wires.[46, 47]

A second mechanism is coherent tunneling between the adjacent sites in an off-resonance regime. In such a case, the decay factor β (Fig. 4) is given by[38, 39]

$$\beta = \frac{2}{\lambda} \ln\left(\frac{\varepsilon_0}{\Gamma_3}\right) \qquad (2)$$

where λ is the site size (here we consider λ = s, the typical size of a PBA nanocrystal). From the β values (Fig. 4) and taking the energy values $\varepsilon_0$ from the I-V fits (Fig. 5), we obtain $\Gamma_3$ = 0.11 – 0.24 and 0.14 – 0.26 eV for CsCoFe and CsNiCr nanocrystals, respectively. Albeit, these values are close to the limit of validity ($\Gamma_3 < \varepsilon_0$) of this model, this result means that we have a strong coupling between the nanocrystals and the 3-nanocrystals junctions behave as a "molecular wire". The inter-particle coupling is almost the same for the two types of objects. With almost the same $\ln(\varepsilon_0/\Gamma_3)$ value, the difference in β (a factor of 2.4 on the mean values) simply reflects the size of PBAs (almost a factor of 2.5). This strong inter-particle coupling energy means that we can consider the nanoscale devices with 2 or 3 particles between the electrodes as an "effective" long one-nanoparticle "wire" with its "effective" electron-transporting orbital at about the energy $\varepsilon_0$. This is consistent with the fact that fitting the single energy level model (Eq. 1) on the average I-V curves of peaks P2 and P3 (see

Figs. S22 and S23 in the supporting information) gives almost the same $\varepsilon_0$ values as for peak P1 (see also the detailed statistic, Figs. S18 and S19 in the Supporting Information).

The PBA nanocrystals are stable in aqueous solution because they are negatively charged,[13, 14] $Cs^+$ ions ensure electric neutrality in solution. Their surface has water molecules coordinated to the metal ions ($Co^{III}$ and $Ni^{II}$ for CsCoFe and CsNiCr, respectively) and nitrogen atoms coming from the hexacyanometalate entities (Scheme 1-b).[33] These objects are, therefore, highly hydrophilic and cannot directly be physisorbed on hydrophobic substrates such as HOPG. However, it has been shown that an ice-like network of water molecules (about 1 nm thickness) forms on HOPG and other hydrophobic substrates when humidity is present.[50-52] Consequently, the presence of a thin layer of water molecules at the interface between the PBAs and the hydrophobic HOPG surface is responsible for a network of hydrogen bonds ensuring their assembly. The same situation may occur at the interface between the AFM tip (PtIr) and the PBAs with the well-known formation of a water meniscus at the tip/surface interface. The presence of such water layer is consistent to explain two features: (i) the relatively weak coupling parameters $\Gamma_1$ and $\Gamma_2$ ($\approx$ 0.5-1.5 meV) between the electrodes and the nanocrystals and (ii) the small difference between $\Gamma_1$ and $\Gamma_2$. The ultra-thin water layer plays the role of a decoupling layer more or less hiding the difference in the nature of the electrodes "seen" by the nanocrystals, in agreement with $\Gamma_1 \approx \Gamma_2$. Typically, the measured coupling parameter in this work (ca. 0.5 - 1.5 meV, see above) is lower than typical values already measured on chemisorbed molecular junctions (ca. 5 to 10 meV), i.e. when molecules are attached on the surface via an anchoring group (e.g. thiol on Au). The presence of such an ultra-thin water layer is consistent with this weak electronic coupling.

The large interparticle coupling parameter ($\Gamma_3 \approx$ 100-250 meV) that is two orders of magnitude larger than $\Gamma_1$ and $\Gamma_2$ is intriguing. It excludes the presence of a water layer and a H-bond network between the nanocrystals as with the electrodes and suggests the occurrence of a direct interparticle contact. Two possible hypotheses can be invoked regarding this contact: (i) either there is coalescence between the nanoobjects located one on top of the other due to the formation of coordination bonds through a substitution of the water molecules linked to the metal ions of one nanoparticle by the nitrogen end of the other one or (ii) an electrostatic attraction between the negatively charged objects occurs through the $Cs^+$ ions present in solution keeping the objects close to each other. The first hypothesis can reasonably be excluded because a genuine orbital overlap (hybridization) between the objects would be present identical to that within the objects themselves and would lead to a loss of their individual electronic character. We, thus, assume that the nanocrystals are separated (in the z direction) by a layer of $Cs^+$ ions, the $\Gamma_3$ coupling parameter of 100-250 meV is consistent with the electrostatic interactions at the interface between two PBA nanocrystals.

The β values are close to those determined for other organometallic wires, in which Co or Fe ions are used to assemble π-conjugated molecules in long wires (up to 40 nm). These authors have reported β ≈ 0.28 $nm^{-1}$ for Fe-based wires and β ≈ $10^{-2}$ $nm^{-1}$ for Co-based wires.[4] They postulated that the electron transport occurs by multi-site hopping via the HOMO of the Co- (or Fe-) metal center molecular wires. Yan et al. reported β values in the range 0.8 $nm^{-1}$ down to 1.5x$10^{-2}$ $nm^{-1}$ for long molecules (8-22 nm) of bis-thienylbenzene derivatives.[48] The same group has measured β ≈ 0.17 – 0.25 $nm^{-1}$ for Co and Ru terpyridine oligomer films (4 – 14 nm thick).[45, 53] Choi et al. have measured β about 0.9 $nm^{-1}$ for conjugated oligophenyleneimine with length in the range 4 - 8 nm.[46] Zhao et al. reported β ≈ 0.16 $nm^{-1}$ for oligo(aryleneethynylene) with length between 3 and 6 nm.[54] These behaviors were associated to hopping transport along the molecule wires.

## Conclusions

In conclusion, we reported electron transport measurements through nanoscale devices consisting of 1 to 3 Prussian blue analog (PBA) nanocrystals connected between a HOPG electrode and the tip of a conducting-AFM. For both types of nanocrystals (CsCoFe and CsNiCr), we observe a long-range electron transport (up to 45 nm), characterized by low decay factors, β ≈ 0.11 - 0.18 $nm^{-1}$ (CsCoFe) and β ≈ 0.25 - 0.34 $nm^{-1}$ (CsNiCr). These decay factors agree with a multi-step coherent tunneling in the off-resonance case between adjacent nanocrystals, with a strong interparticle coupling (around 0.1 – 0.25 eV), due to electrostatic interactions. The electron transport in single PBA junction is experimentally determined (from our current-voltage measurements) at around 0.5 eV from the electrode Fermi energy in the two cases. From a comparison with the calculated electronic structures of the PBAs, we identify that, the electron transport is mediated by the localized d bands and we suggest that the involved orbitals are the filled $Fe^{II}$-$t_{2g}$ d band (HOMO) for CsCoFe and the half-filled $Ni^{II}$-$e_g$ d band (SOMO) for CsNiCr. The decay factor β for the CsCoFe nanocrystals (45 nm) is, to date, almost one order of magnitude weaker than those of MOFs measured in similar conditions.[1, 55] It is close to the best values obtained for metal-containing molecular wires.[4, 5, 45, 48, 53, 56, 57] PBA nanocrystals are, therefore, competitive in terms of relaying electrons over relatively long distances. The values measured here for 45 and 18 nm distances between the electrodes correspond to a stack of three nanocrystals in close contact and not to a continuous one nanocrystal of 45 and 18 nm thicknesses. Even with a strong coupling between adjacent nanocrystals, it is expected that a single nanocrystal with the same length would have better performance. We have already demonstrated that we can control the size of these objects from 6 and up 80 nm keeping them stable in solution,[14, 29] which may allow their assembly with increasing size on HOPG to measure their transport behavior. The intrinsic conductance of the nanocrystals depends, among other things, on the position of the HOMO (SOMO) and LUMO bands with respect to the Fermi levels of the electrodes, which is determined by the electronic configuration of the metal ions and the degree of overlap between their d orbitals and the p orbitals of the bridging cyanide ligand. The better the overlap is, the wider the HOMO and LUMO bands are, leading to a lower energy mismatch with the electrode Fermi levels. Thanks to the chemical versatility of the PBAs and other related cyanide-bridged networks, it is possible to access almost any combination of metal ions including those belonging to second and third row transition metal where the d-p overlap is expected to be large. Work on such systems possessing different sizes is underway.

**Methods.**

*Sample fabrication.*

The CsCoFe and CsNiCr PBAs were synthesized as reported elsewhere[13] and briefly summarized in the Supporting Information. Their assembly on HOPG was carried out as follow: a freshly prepared aqueous colloidal dispersion of the nanoobjects was prepared. A highly oriented pyrolytic graphite (HOPG) substrate was cleaved by a scotch tape and immediately immerged in the colloidal solution. The vial temperature was maintained at room temperature for the CsCoFe dispersion and at 4 °C for CsNiCr. The substrates were removed from solution after 20 s of immersion time and rinsed thoroughly with water and then with methanol for CsCoFe and only with methanol for CsNiCr and then both dried under vacuum for several hours.

*Physico-chemical characterizations.*

The dynamic light scattering measurements were performed on a Malvern Nanozetasizer Apparatus (equipped with a backscattering mode) on the aqueous solutions (1.5 mL) containing the particles. The volume profile was used to estimate the size corresponding to the main peaks. This measurement was used as a qualitative measurement of the size of the particles or aggregates in solution, which systematically includes a solvation shell.

The TEM measurements were done on a TEM Philips EM208 with 100 keV incident electrons focused on the specimen.

Powder X-ray diffraction (XRD) was performed on powders deposited on an aluminum plate and collected on a Philipps Panalytical X'Pert Pro MPD powder diffractometer at CuKα radiation equipped with a fast detector.

FT-IR spectra were recorded with a Perkin Elmer spectrometer (Spectrum 100). The measurements were performed on KBr pellets (typically 1 mg in ca. 99 mg of KBr, this latter being previously ground) in the 300–4000 cm$^{-1}$ range.

EDS were performed with a ZEISS FEG-SEM Supra 55 VP with an electron beam at 15 kV. The samples were imaged by scanning electron microscopy (ZEISS ULTRA 55, at 7kV, secondary electron images) and atomic force microscopy (Innova, Bruker) in the tapping mode (TM). AFM images and histograms of heights and particle sizes were processed with Gwyddion software.[58] The SEM images were analyzed with ImageJ software (imagej.nih.gov).

*Electrical characterization*.

The electron transport properties at the nanoscale were measured by C-AFM (ICON, Bruker) at room temperature under a flow of dry nitrogen using a tip probe in platinum/iridium. We used a low tip loading force of ca. 3 nN to avoid a too important strain-induced deformation of the molecular film (≲ 0.3 nm).[35] In addition, we did not record scanning current images (contact mode) to avoid any distortion of the PBA structures deposited on the HOPG. We used a "blind" mode to measure the current-voltage (I-V) curves and the current histograms: a square grid of 10×10 was defined with a pitch of 50 nm. At each point, the I-V curve is acquired leading to the measurements of 100 traces per grid. This process was repeated several times at different places (randomly chosen) on the sample, and up to thousands of I-V traces were used to construct the current-voltage histograms. Two sets of I-V curves were measured with two sensitivities of the current preamplifier (PFTUNA, ICON, Bruker), first at a sensitivity of 1nA/V, and then at 20 pA/V for a better resolution of the currents below $10^{-10}$ A. The voltage was applied on the HOPG substrate, the C-AFM tip grounded. The fits of Eq. 1 were performed with the routine included in ORIGIN software, using the method of least squares and the Levenberg Marquardt iteration algorithm. All fits showed $R^2 > 0.98$ with intrinsic parameter errors of less than 0.02 eV for $\varepsilon_0$ and less than 0.03 meV for $\Gamma_1$ and $\Gamma_2$.

## Supporting information.

The Supporting Information is available free of charge at xxxxxx.

Preparation and full characterization of the CsNiCr and CsCoFe nanocrystals. Additional AFM images and histograms of heights of the PBA layers. Analysis of the particle size on surfaces (AFM and SEM images). Current-voltage measurements on the bare HOPG substrates. Histograms of the tip z-position during the C-AFM measurements. Topographic AFM and conducting AFM on thick films. Detailed statistical analysis of the decay factor β. Additional fits of the single molecular energy level model on peak P1. Detailed statistical analysis of the energy levels. Conductance versus 1/distance and electric field. Fits of the single molecular energy level model on peaks P2 and P3.

## Author contributions.

R.B. carried out the C-AFM, TM-AFM measurements with the help of S.L., who also performed the SEM measurements. S.M. and T.M. synthesized the nanocrystals, performed the deposition on HOPG and carried out the preliminary AFM imaging. T.M. and D.V. conceived and supervised the project. D.V. analyzed the experimental data, carried out the theoretical analysis and wrote the paper with contributions from all authors. All authors have given approval to the final version of the manuscript.

## ORCID.

R.Bonnet: 0000-0001-7327-2603
T. Mallah: 0000-0002-9311-3463
S. Lenfant: 0000-0002-6857-8752
D. Vuillaume: 0000-0002-3362-1669

## Conflict of interest.

The authors declare no competing financial interests.

## Acknowledgements.

This work was financially supported by an ANR grant "Spinfun" (n° ANR-17-CE24-0004).

# Long-range electron transport in Prussian blue analog nanocrystals.

Roméo Bonnet[1,*], Stéphane Lenfant[1], Sandra Mazérat[2],
Talal Mallah[2,#] and Dominique Vuillaume[1,#]

1. Institute for Electronics Microelectronics and Nanotechnology (IEMN), CNRS, Av. Poincaré, 59652 Villeneuve d'Ascq, France.

2. Institut de Chimie Moléculaire et des Matériaux d'Orsay (ICMMO), CNRS, University Paris-Sud, Rue du doyen Georges Poitou, 91405 Orsay, France.

* Now at: ITODYS, CNRS, Univ. Paris-Diderot, 15, rue Jean Antoine de Baïf, 75013 Paris, France.

# Corresponding authors : talal.mallah@u-psud.fr; dominique.vuillaume@iemn.fr

## SUPPORTING INFORMATION

1. Preparation and full characterization of the CsNiCr and CsCoFe nanoparticles

2. Additional AFM images and histograms of heights of the PBA layers.

3. Analysis of the particle size on surfaces (AFM and SEM images).

4. Histograms of the tip z-position during the C-AFM measurements.

5. Current-voltage measurements on the bare HOPG substrates.

6. Topographic AFM and conducting AFM on thick films.

7. Detailed statistical analysis of the decay factor β.

8. Additional fits of the single molecular energy level model on peak P1.

9. Detailed statistical analysis of the energy levels.

10. Conductance versus 1/distance and electric field.

11. Fits of the single molecular energy level model on peaks P2 and P3.

# 1. Preparation and characterization of the CsNiCr(CN)$_6$ (abbreviated as CsNiCr) and CsCoFe(CN)$_6$ (abbreviated as CsCoFe) nanocrystals

*CsNiCr nanocrystals.*

1 - The nanocrystals were prepared as in ref. 1: an aqueous solution (100 mL) containing NiCl$_2$.6H$_2$O (0.2 x10$^{-3}$ mol, $c$ = 2x10$^{-3}$ M) and CsCl (0.4x10$^{-3}$ mol, c = 4x10$^{-3}$ M) was added rapidly in an equal volume of an aqueous solution of K$_3$[Cr(CN)$_6$] (0.2x10$^{-3}$ mol, $c$ = 2x10$^{-3}$ M) under vigorous stirring. The solution was stirred for one hour.

2 - Dynamic light scattering (DLS) of the as-prepared solution was measured and shows a hydrodynamic diameter of 6 nm (Figure S1). The zeta potential measured was found to be equal to -31 mV indicating the presence of negatively charged particles in the solution.

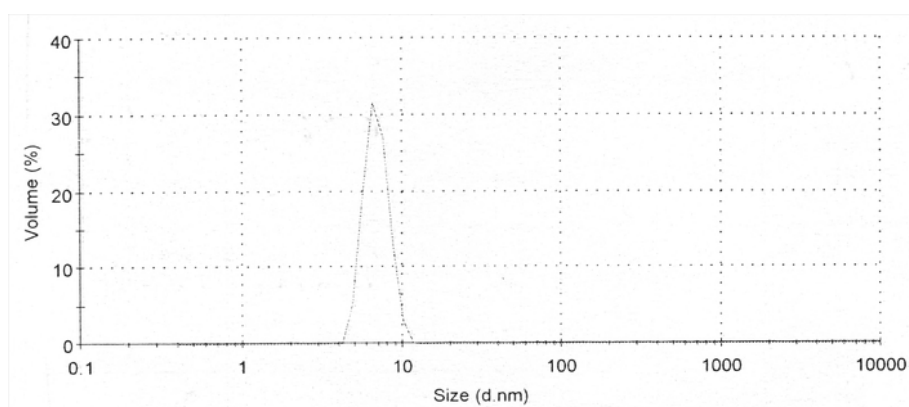

**Figure S1**. DLS of the as-prepared colloidal solution of CsNiCr nanocrystals.

3 - Transmission Electron Microscopy (TEM) imaging of the as-prepared nanocrystals show cubic objects with a size close to 6 nm confirming the DLS data. It is worth noting that the objects are not very stable under the electron beam making difficult to focus on small areas. TEM images were acquired on different areas of the grid giving the same results (Figure S2).

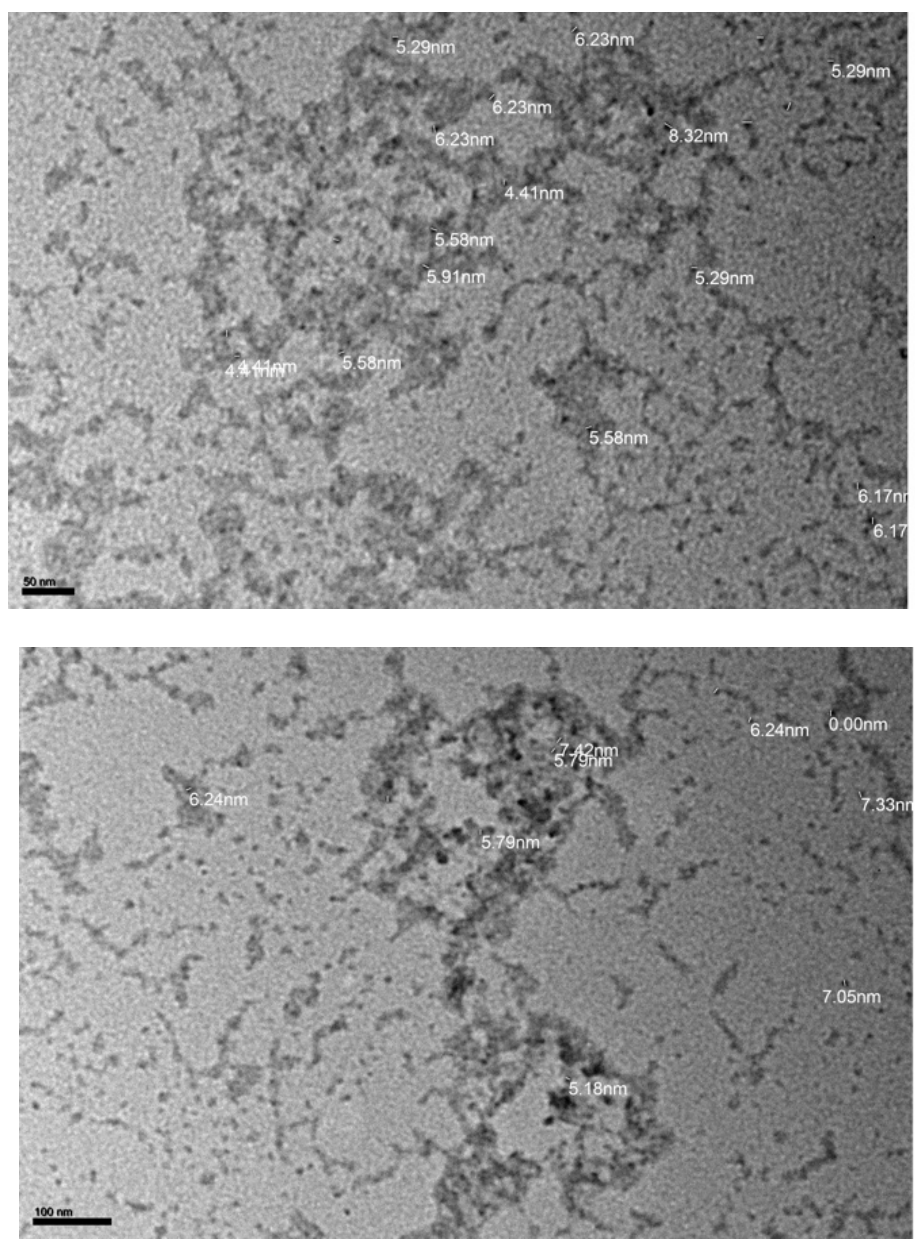

**Figure S2**. Transmission Electron Microscopy imaging of the CsNiCr nanocrystals on two different regions of the grid.

4 - Infra-red spectroscopy of the nanocrystals was performed on a solid material obtained by recovering the objects after adding excess methanol on the colloidal solution. The spectrum (Figure S3) show in the 2200-2000 cm$^{-1}$ region the asymmetric vibration mode of the cyanide at 2171 cm$^{-1}$ characteristics of cyanide bridging a trivalent metal ion (Cr(III) here) and a divalent one (Ni(II) here) corresponding to the Cr-CN-Ni sequence as expected.[1] The shoulder at 2134 cm$^{-1}$ is assigned to non-bridged cyanides coordinated to Cr(III) as expected for $Cr(CN)_6$ species present at the surface of the nanocrystals.

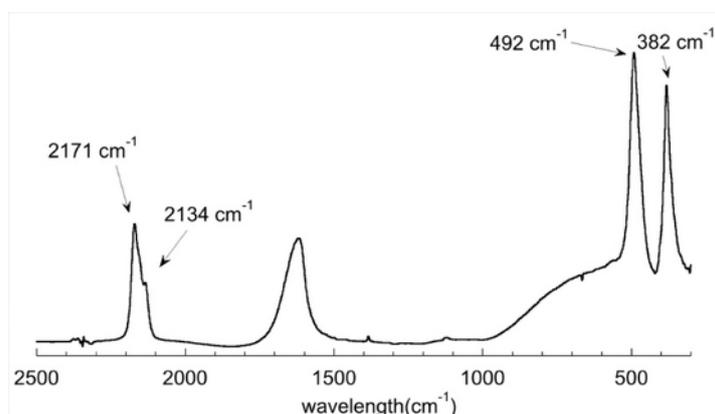

**Figure S3**. Infra-red spectrum of CsNiCr nanocrystals, the band at 1630 $cm^{-1}$ corresponds to water molecules.

5 - X-ray diffraction on a powder sample obtained by recovering the nanocrystals by addition of an excess of CTABr (Cetyltrimethylammonium bromide) show a pattern (Figure S4) corresponding to a face centered cubic structure with a cell parameter $a$ = 10.50 Å as expected.[1] The size of the crystalline domains determined using the Scherrer equation was found equal to 6.2 nm confirming the DLS and TEM data.

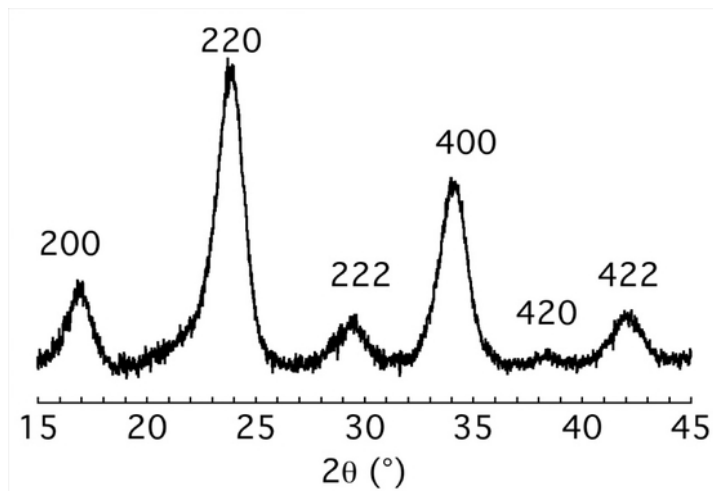

**Figure S4**. Powder X-ray diffraction pattern of the CsNiCr nanocrystals.

6 - Energy dispersive X-ray Spectroscopy carried of the as-prepared nanocrystals casted on a grid give the following atomic percentages Cs (33.17%), Ni (34.36%) and Cr (32.46%) corresponding to the following formula: $Cs_{0.96}Ni[Cr(CN)_6]_{0.94}$.

*CsFeCo nanocrystals.*

1 - The CsFeCo(CN)$_6$ nanocrystals were prepared using the same procedure as for the CsNiCr(CN)$_6$ ones replacing Ni(II) and Co(II) and Cr(III) by Fe(III).

2 - Dynamic light scattering (DLS) of the as prepared colloidal solution (Figure S5) show the presence of objects with hydrodynamic diameter of 15 nm.

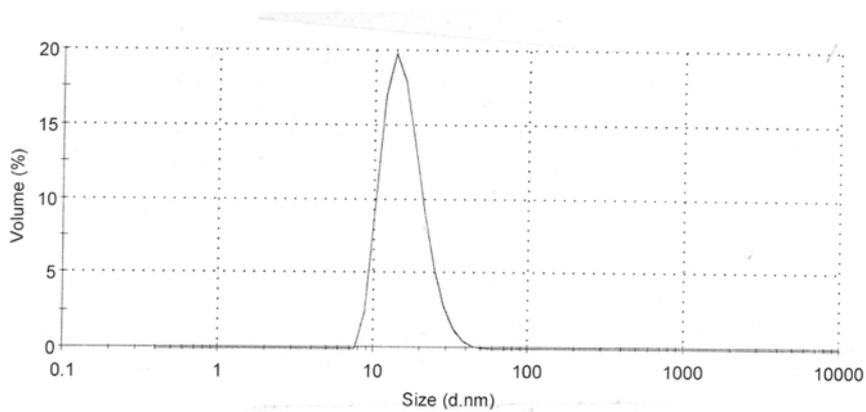

**Figure S5**. DLS of the as-prepared colloidal solution of CsCoFe nanocrystals.

3 - Transmission Electron Microscopy (TEM) imaging of the as-prepared nanocrystals show cubic objects with a size close to 15 nm confirming the DLS data.

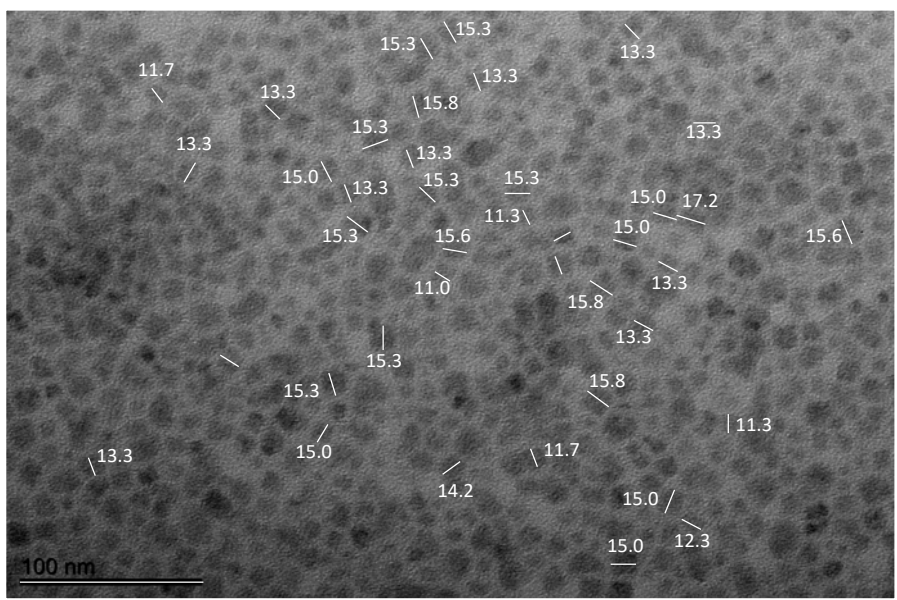

**Figure S6**. Transmission Electron Microscopy imaging of the CsCoFe nanocrystals.

4 - Infra-red spectroscopy of the nanocrystals was performed on a solid material obtained by recovering the objects after precipitating the colloidal solution with $CaCl_2$. The spectrum (Figure S7) show in the 2200-2000 $cm^{-1}$ region the asymmetric vibration mode of the cyanide at 2120 $cm^{-1}$ characteristics of cyanide bridge corresponding to the Fe(II)-CN-Co(III) sequence as expected for a when an electron transfer occurs during the reaction between $Fe^{III}(CN)_6^{3-}$ and $Co^{II}(H_2O)_6^{2+}$.[2]

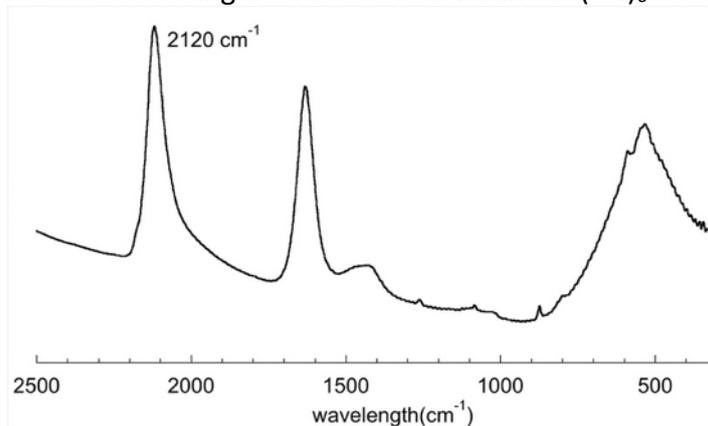

**Figure S7**. Infra-red spectrum of CsCoFe nanocrystals, the band at 1630 $cm^{-1}$ corresponds to water molecules.

5 - X-ray diffraction on a powder sample obtained by recovering the nanocrystals by addition of an excess of CTABr (Cetyltrimethylammonium bromide) show a pattern (Figure S8) corresponding to a face centered cubic structure with a cell parameter $a$ = 10.02 Å as expected for the presence of a majority of $Fe^{II}$-CN-$Co^{III}$ pairs within the nanocrystals.[3] The size of the crystalline domains determined using the Scherrer equation was found equal to around 11 nm.

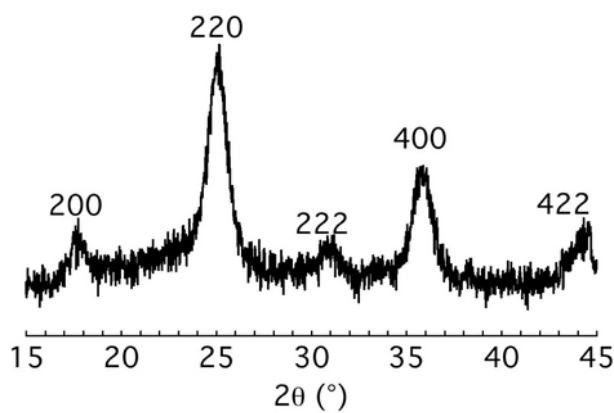

Figure S8. Powder X-ray diffraction of the CsFeCo nanocrystals.

6 - Energy dispersive X-ray Spectroscopy carried of the nanocrystals recovered by addition of an excess of CTABr (Cetyltrimethylammonium bromide), redispersed in methanol and casted on a grid give the following atomic percentages Cs (27.61%), Co (38.7%) and Fe (33.7%) corresponding to the following formula: $Cs_{0.7}Co[Fe(CN)_6]_{0.9}$. This formula does not include the $CTA^+$ that acts as counter-anions for the negatively charged nanocrystals, the Cs$^+$ ions are present within the tetrahedral sites of the fcc nanocrystals.

**2. Additional AFM images and histograms of heights.**

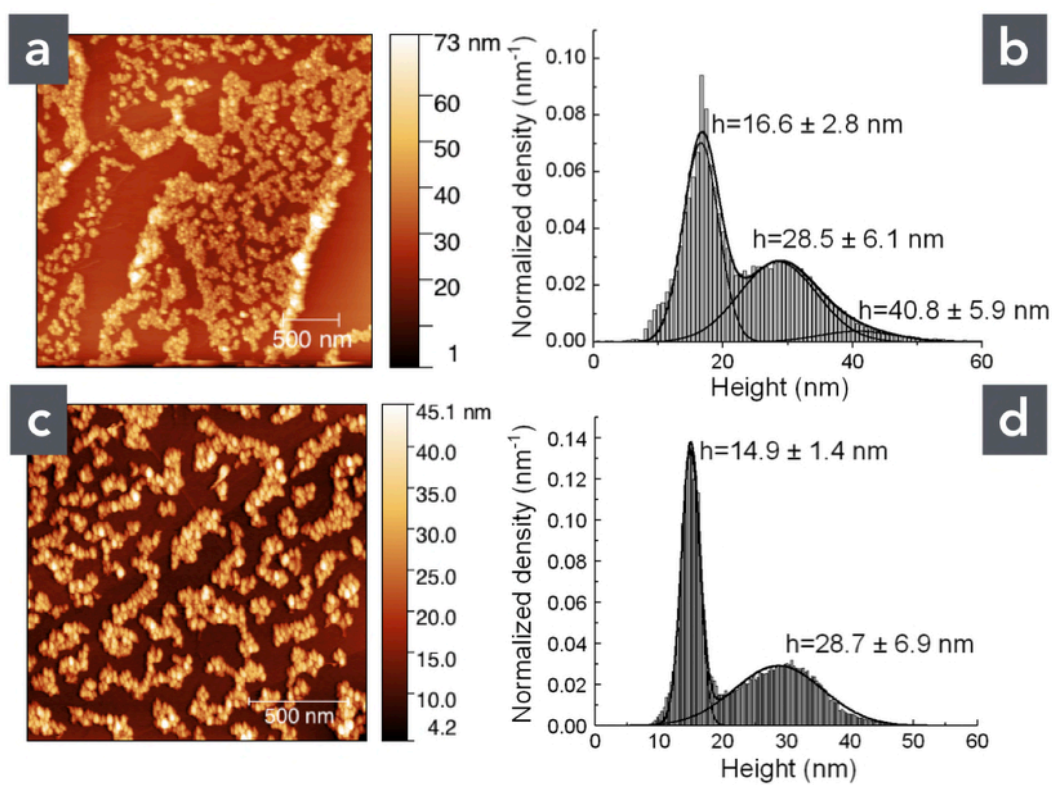

**Figure S9. (a, c)** Topographic AFM image and corresponding histograms of heights **(b, d)** of the CsCoFe layers. Fig. S9-a is a zoom of the Figure 1-c in the main text. Fig. S9-c is taken on another zone of the sample. In panels (b) and (d), the values marked for each peaks are the average heights and the standard deviations obtained from the fits of multi-Gaussian distributions (black lines).

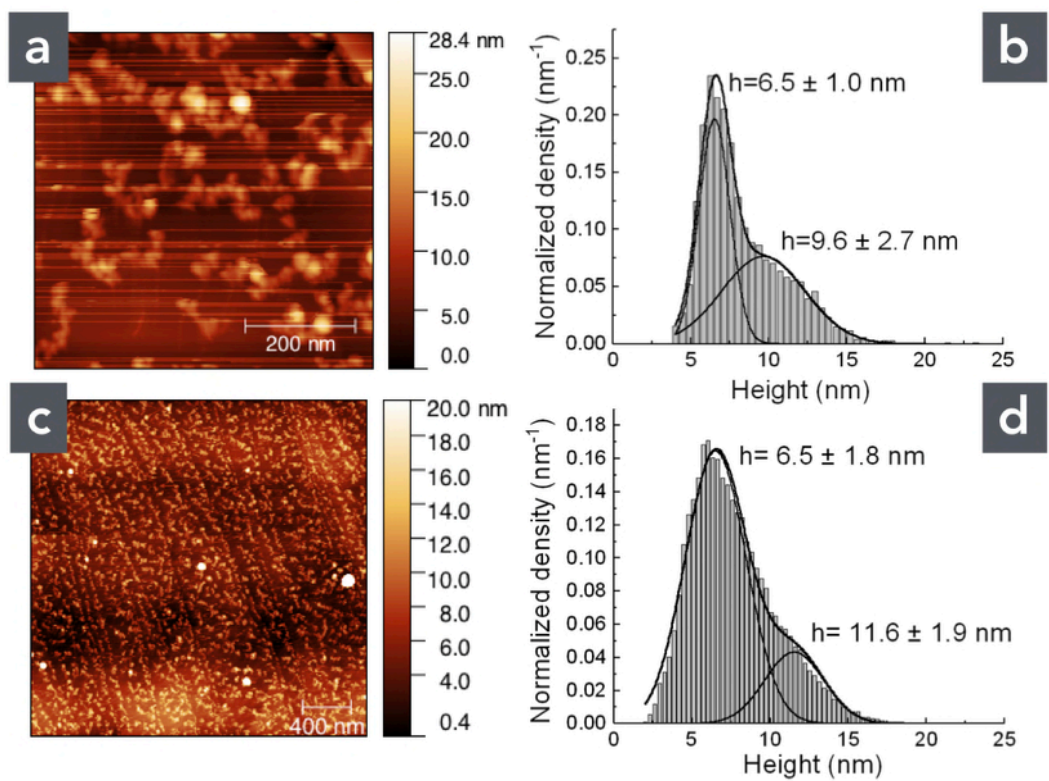

**Figure S10. (a, c)** Topographic AFM image and corresponding histograms of heights **(b, d)** of the CsNiCr layers taken on two other zones of the sample. In panels (b) and (d), the values marked for each peaks are the average heights and the standard deviations obtained from the fits of multi-Gaussian distributions (black lines). In (a, b) the horizontal defects/lines (image instabilities) have been masked for the histogram analysis.

## 3. Analysis of the particle size on surfaces (AFM and SEM images).

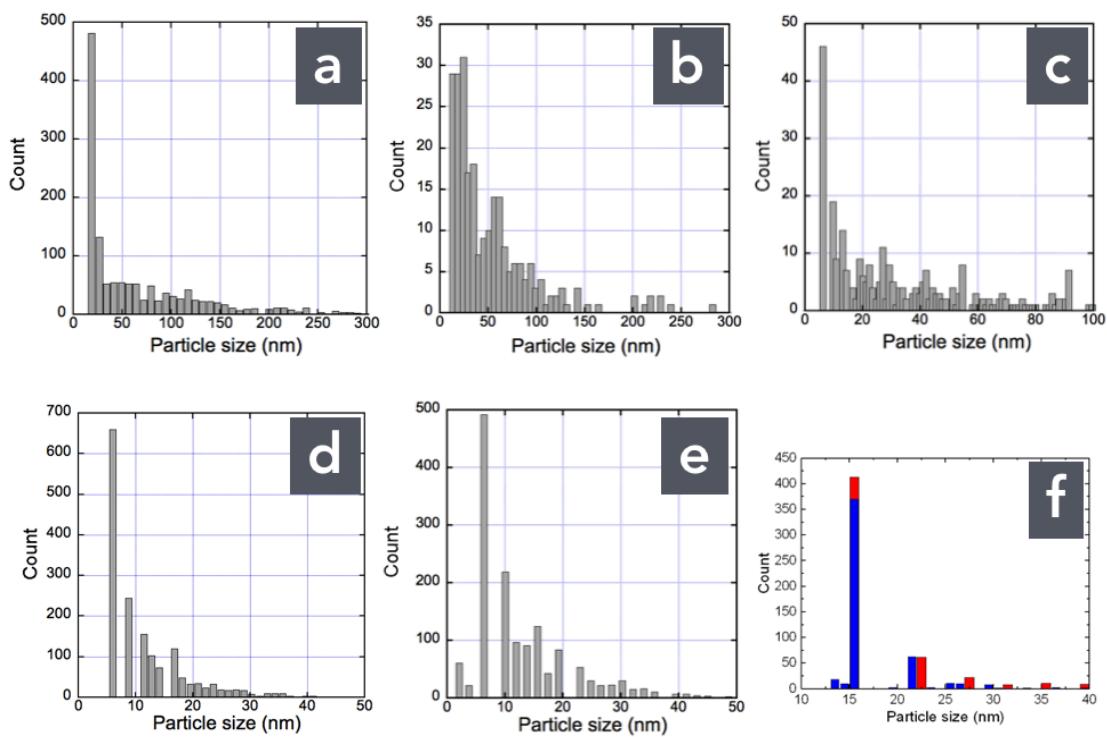

**Figure S11. (a,b,c)** CsCoFe : histograms of the particle sizes calculated from the AFM images Figs. 1-c, S9-a and S9-c, respectively. **(d,e)** CsNiCr : histograms of the particle sizes calculated from the AFM images Figs. 1-e and S10-c, respectively. The thresholding method was used for the grain analysis, and, assuming a square face of the PBA, the plotted particle size is the square root of the calculated projected (flat) area of the grain. **(f)** CsCoFe : histograms of the particle sizes calculated from the SEM images Figs. 1-a (in red) and Fig. 1-b (in blue).

## 4. Histograms of the tip z-position during the C-AFM measurements.

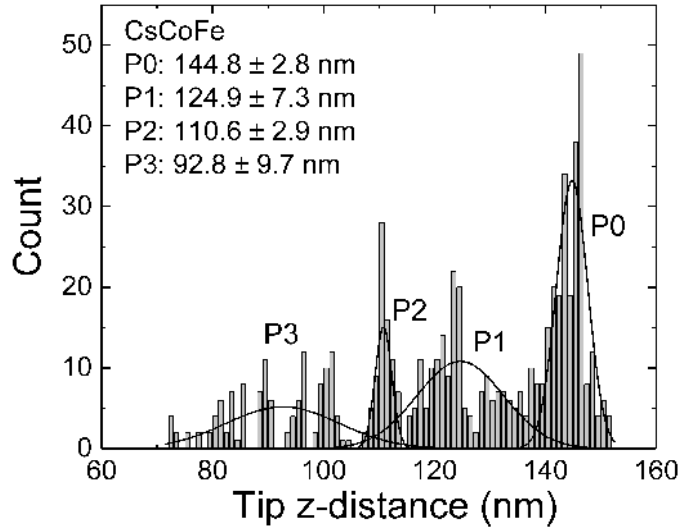

**Figure S12.** CsCoFe. Histograms of the tip z-position (0 = tip retracted). The z-position is recorded simultaneously for each I-V curve (500 I-V traces). The mean value and standard deviation (Gaussian fits) are given in the figure for each peaks.

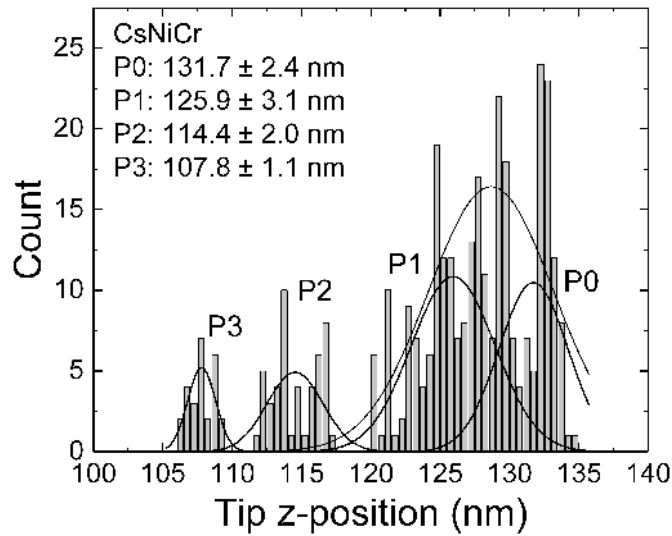

**Figure S13.** CsNiCr. Histograms of the tip z-position (0 = tip retracted). The z-position is recorded simultaneously for each I-V curve (500 I-V traces). The mean value and standard deviation (Gaussian fits) are given in the figure for each peaks. The large peak (thin line) can be decomposed into 2 contributions (P0 and P1).

## 5. Current-voltage measurements on the bare HOPG substrates.

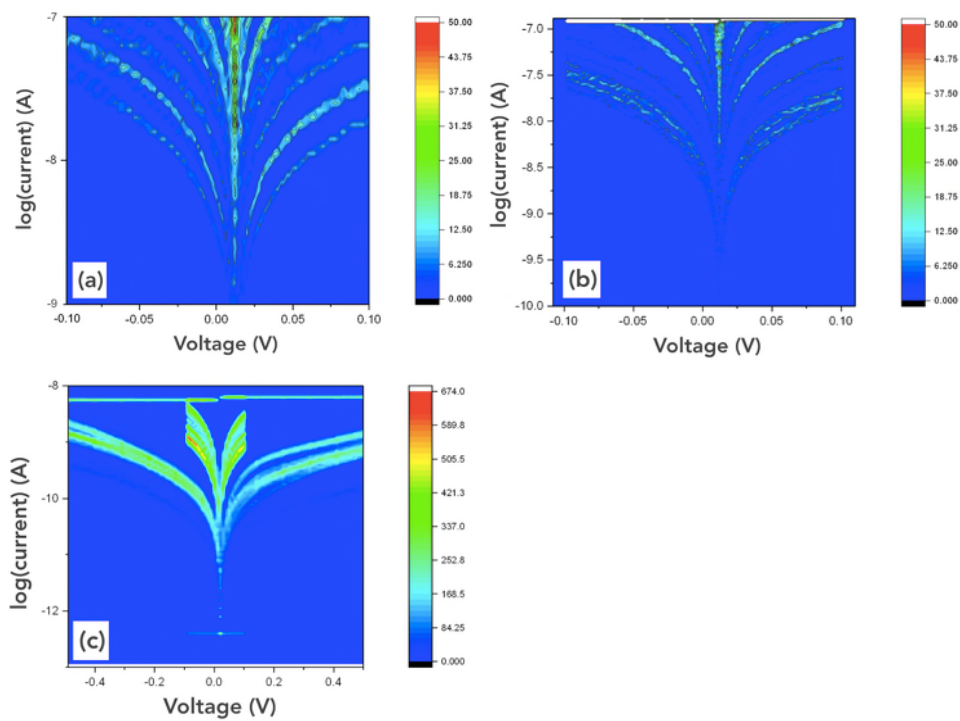

**Figure S14. (a and b)** Typical I-V 2D histograms of the bare HOPG substrates used for the deposition of CsCoFe (2 zones). The I-V curves correspond to peak P0 in Fig. 2-a (currents above 10 nA at low voltages ±0.1 V). **(c)** Typical I-V 2D histograms of the bare HOPG substrates used for the deposition of CsNiCr showing three zones (currents above current preamp saturation - horizontal line, currents between 1 and 10 nA at ±0.1V and a zone less conducting with currents around nA at ± 0.5 V corresponding to peak P0 in Fig. 3-a. All currents plotted on decimal log scales.

## 6. Topographic AFM and conducting AFM on thick films.

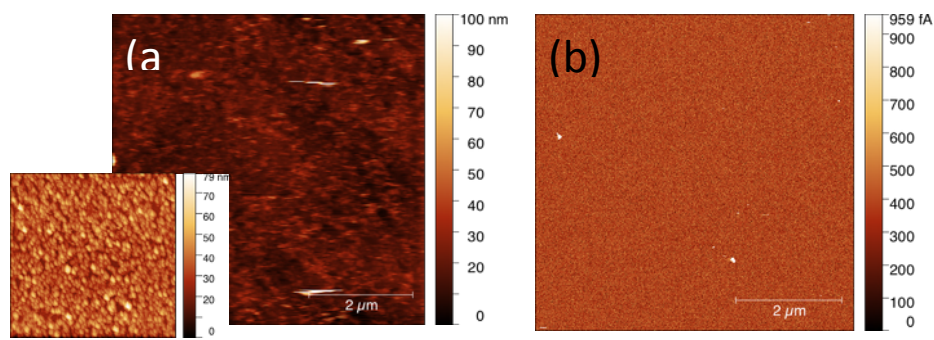

**Figure S15. (a)** Topographic AFM image of a thicker CsCoFe film with a complete coverage of the HOPG substrate (inset: zoom on a 2 μm x 2 μm zone), and **(b)** corresponding conducting-AFM (at 50 mV) showing no current (below the sensitivity limit of about $5 \times 10^{-13}$ A).

## 7. Detailed statistical analysis of the decay factor β.

For each I-V curve in the data sets of Figs. 2 and 3, we plot the current measured at 50, 200 and 400 mV versus the tip height relative to the surface. This value is deduced from the z-position, z, which is recorded simultaneously for each I-V traces (Figs. S12 and S13) and the height, h, relative to the surface, PBA layer thickness, is calculated by h = -(z - z(P0)), where z(P0) is the average z value for the peak P0 determined in the histograms in Figs. S12 and S13. We discard from this analysis the I-V traces that are too noisy (especially for low current below $10^{-12}$ A) or exhibit a too large hysteresis during the back and forth voltage sweep or abrupt "staircase-like" jumps in the current, which may indicate instabilities of the C-AFM tip contact on the samples. The resulting data (Fig. S16) are fitted by a linear regression to deduce the decay factor β indicated on the plots.

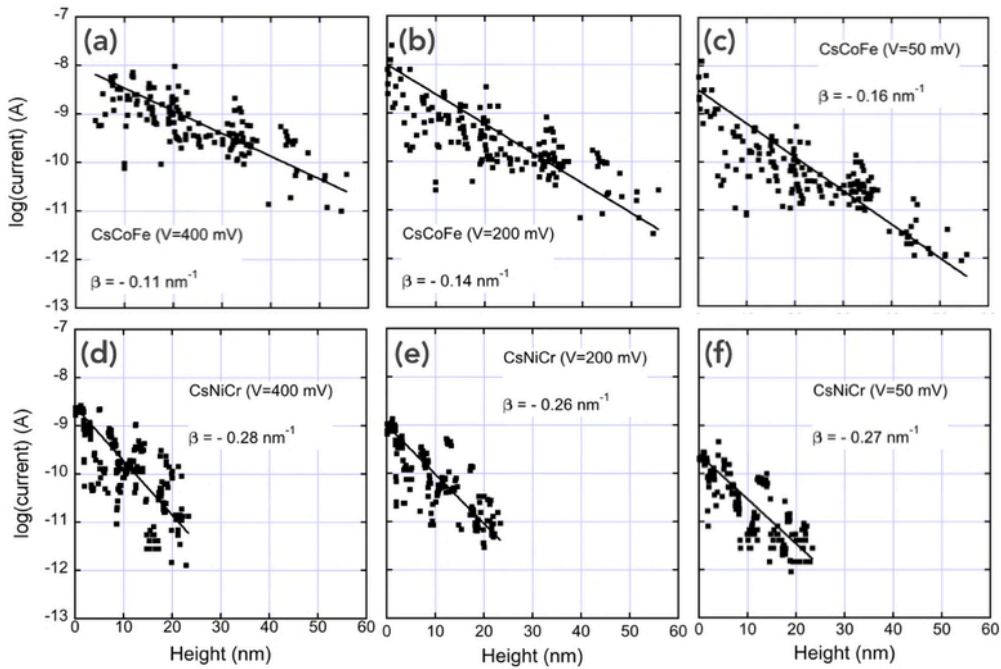

**Figure S16**. Decimal log of current (at V=400, 200 and 50 mV) versus PBA layer height for the CsCoFe, 354 I-V traces **(a-c)** and CsNiCr, 368 I-V traves **(d-e)** samples.

**8. Additional fits of the single molecular energy level model on peak P1.**

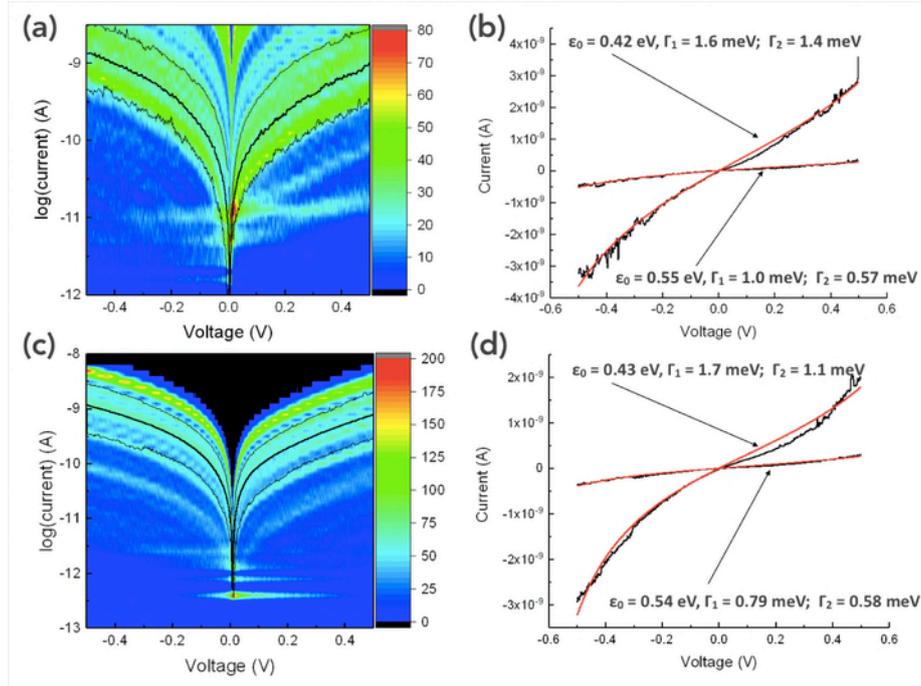

**Figure S17. (a)** CsCoFe. The thin black lines are the IV curves bounding the Gaussian distribution of peak P1 (the bold black line is the average IV, see Fig. 2-a). **(b)** Fits (red lines) of the single molecular energy level model (Eq. 1, main texte) on the two lower and upper limit current voltage curves (black lines) shown in Fig. S2-a. The fitted parameters are given on each figures. **(c and d)** Same as (a and b) for peak P1 of CsNiCr. (panels a and c, currents plotted on decimal log scale).

**9. Detailed statistical analysis of the energy levels.**

Following the same protocol as for Figs. S16, we fit the single energy level model (Eq. 1 in main text) on each I-V traces from the sets of data of the CsCoFe and CsNiCr samples. We discard the I-V traces that are too noisy (especially for low current below $10^{-12}$ A) or exhibit a too large hysteresis during the back and forth voltage sweep or abrupt "staircase-like" jump in the current, since the fits are not accurate enough or significant in such cases. Figs. S18-a and S19-a show the $\varepsilon_0$ values versus height (as determined above, section 7) for the CsCoFe and CsNiCr samples, respectively. Figs. S18-c and S19-c show the histograms of the $\varepsilon_0$ values and the Gaussian fits. In both figure, the panels "b" plot the histograms of heights. The dashed lines are simulated Gaussian curves with the height parameters (mean values and standard deviations) directly determined from the topographic (TM-AFM) images shown in Fig. 1 (main text). Albeit the determination of the heights used for these statistical analyses (also for Figs. S16) is indirect and likely less accurate than a direct height measurement as in the topographic TM-AFM, the agreement is satisfactory. We also observe peaks corresponding of the configurations with one (P1), two (P2) and three (P3) PBA nanocrystals in the HOPG/PBAs/C-AFL tip nanodevices. We just note a weak shift of the peak maxima between the two approaches (about 2 nm), which remains reasonable given the weaker accuracy of the indirect height determination.

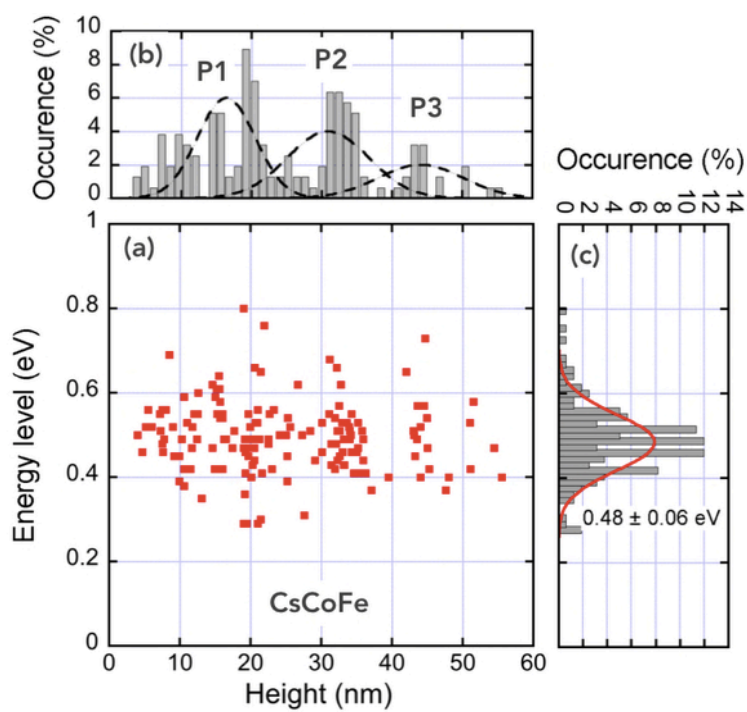

**Figure S18**. CsCoFe. **(a)** Statistical determination of the energy value $\varepsilon_0$ (from 354 I-V traces), and corresponding **(b)** histogram of heights and **(c)** histograms of energy values. The red line is a Gaussian fit given $\varepsilon_0$ = 0.48 ± 0.06 eV.

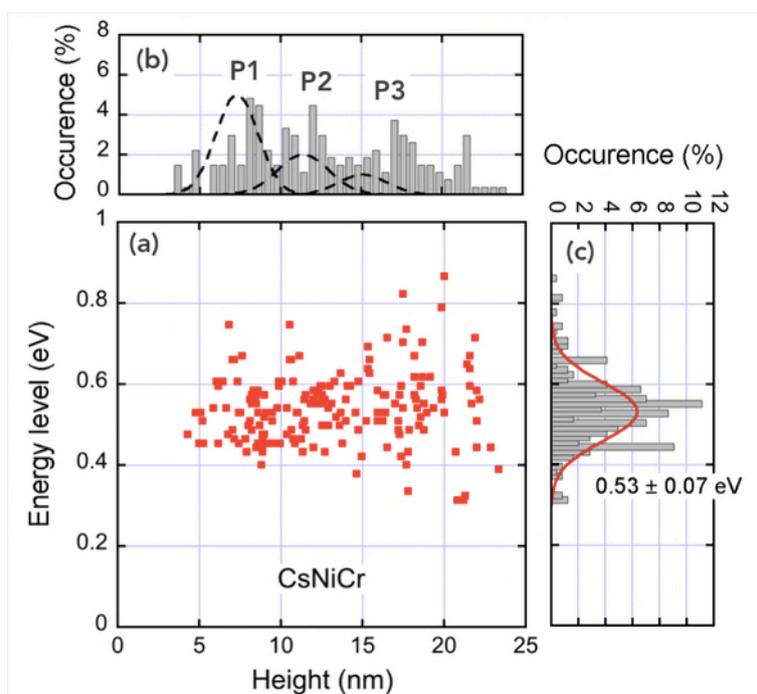

**Figure S19**. CsNiCr. **(a)** Statistical determination of the energy value $\varepsilon_0$ (from 368 I-V traces), and corresponding **(b)** histogram of heights and **(c)** histograms of energy values. The red line is a Gaussian fit given $\varepsilon_0$ = 0.53 ± 0.07 eV.

**10. Conductance versus 1/distance and electric field.**

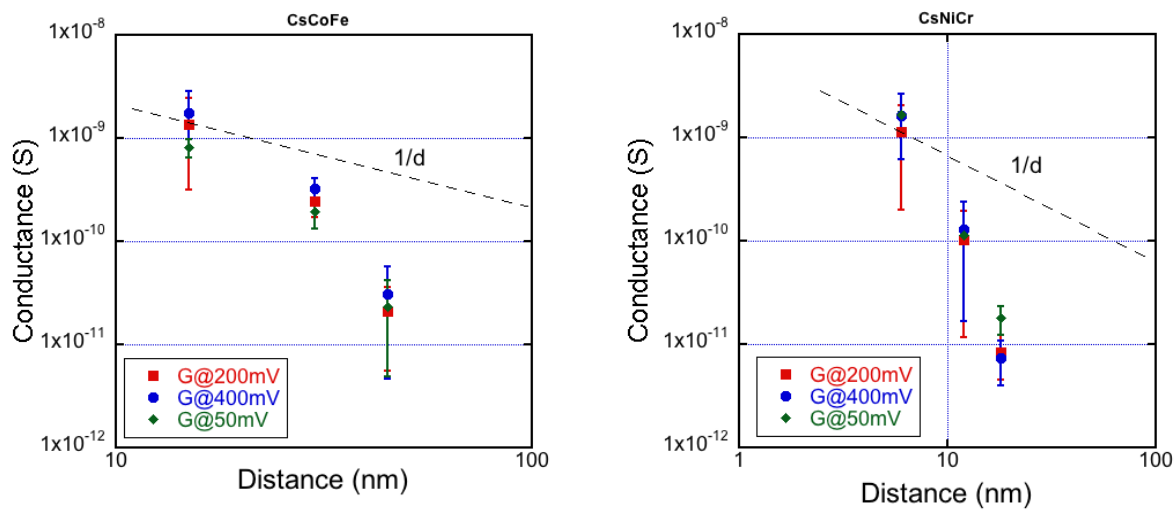

**Figure S20.** Conductance (calculated from the average current peaks P1,P2 and P3 divided by the applied voltage : blue points at 400 mV, red points at 200 mV, green points at 50 mV) vs. 1/d (log-log scale, expected 1/d behavior shown by dashed lines with a slope -1).

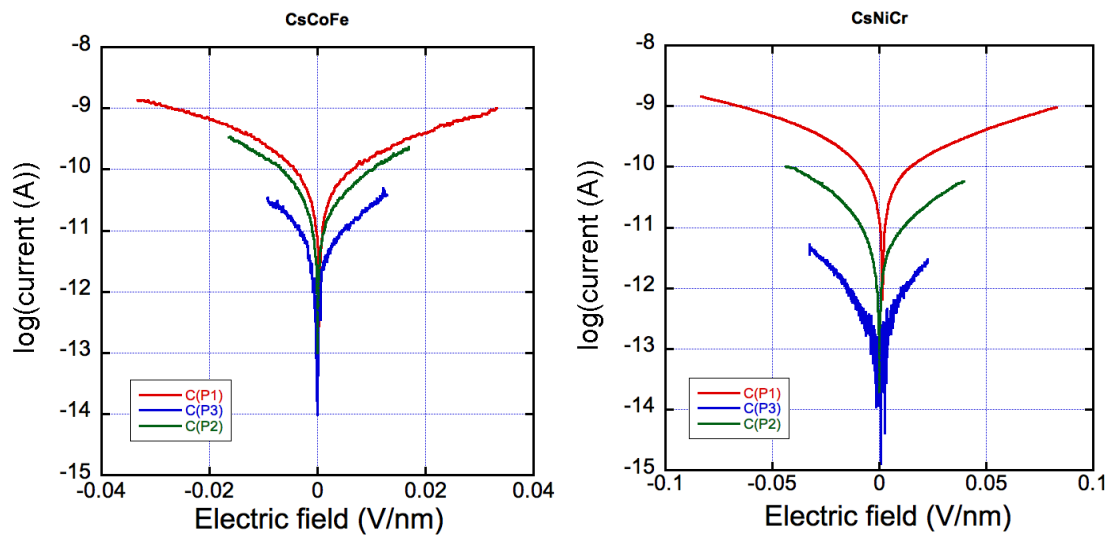

**Figure S21.** Average I-V for the 3 peaks (P1 in red, P2 in green, P3 in blue, all currents plotted on decimal log scales) vs. electric field. They are not superimposed as it should be for a field-driven hopping mechanism.[4, 5]

**11. Fits of the single molecular energy level model on peaks P2 and P3.**

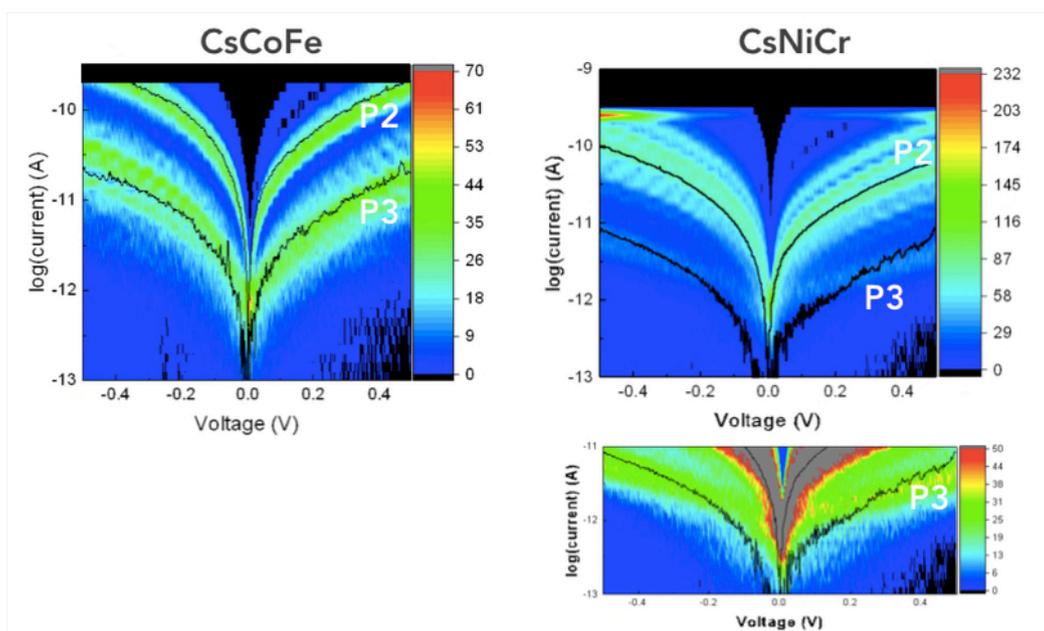

**Figure S22.** Average I-V curves (all currents plotted on decimal log scales) for the peaks P2 and P3 (218 I-V traces for P2 CsCoFe; 123 I-V traces for P3 CsCoFe; 353 I-V traces for P2 CsNiCr). Note that the average I-V for the CsNiCr peak P3 is not very accurate since data are very noisy at this low level

of current and we have only a small number of I-V traces (<50) for this peak (lower right graph is a zoom on P3).

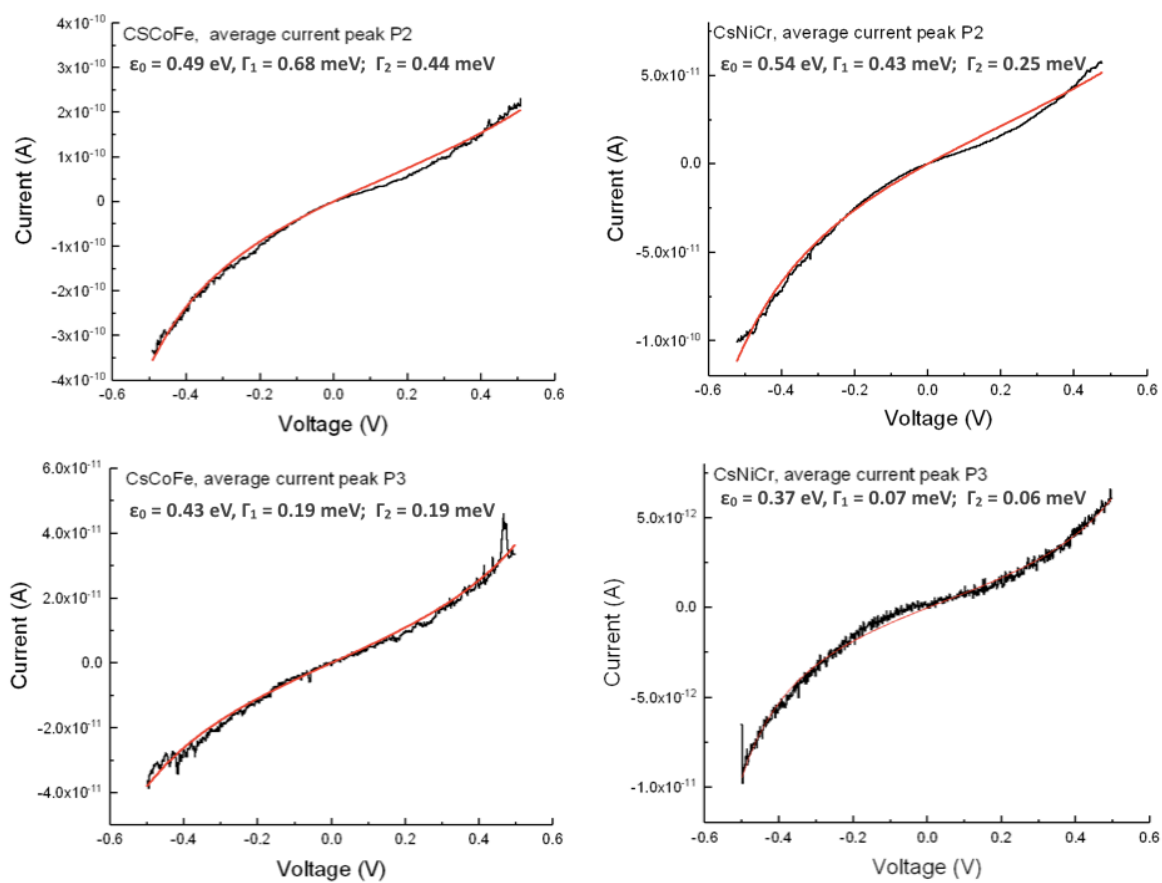

**Figure S23.** Fits (red lines) of the single molecular energy level model (Eq. 1, main texte) on the average current voltage curves (black lines) for the peaks P2 and P3. The fitted parameters are given on each figures. We note a decrease of the $\varepsilon_0$ value for the CsNiCr peak P3 (with respect of the values for P1 and P2), which may be due to the inaccurate determination of the average I-V (see figure S19).